\documentclass[journal=jacsat,manuscript=article]{achemso}

\usepackage[version=3]{mhchem} 
\setkeys{acs}{keywords = true}
\usepackage{graphicx}
\graphicspath{{images/}}

\usepackage{subcaption}
\usepackage{dcolumn}
\usepackage{tabularx,booktabs}
\usepackage{adjustbox} 
\usepackage{bm}
\usepackage{xcolor}
\usepackage{color, colortbl}
\usepackage{ulem}
\definecolor{Gray}{gray}{0.9}

\title{Prediction of Frequency-Dependent Optical Spectrum for Solid Materials: A Multi-Output \& Multi-Fidelity Machine Learning Approach}

\author{Akram Ibrahim}
\affiliation{%
Department of Physics, University of Maryland Baltimore County, 1000 Hilltop Circle, Baltimore, MD 21250, USA
}%
 
\author{Can Ataca}
\affiliation{%
Department of Physics, University of Maryland Baltimore County, 1000 Hilltop Circle, Baltimore, MD 21250, USA
}%

\email{ataca@umbc.edu}
\date{\today}

\keywords{Graph neural networks, Transfer learning, Fidelity embedding, Dielectric function, Absorption coefficient, Solar cells}

\begin{document}

\begin{abstract}
The frequency-dependent optical spectrum is pivotal for a broad range of applications, from material characterization to optoelectronics and energy harvesting. Data-driven surrogate models, trained on density functional theory (DFT) data, have effectively alleviated the scalability limitations of DFT while preserving its chemical accuracy, expediting material discovery. However, prevailing machine learning (ML) efforts often focus on scalar properties such as the band gap, overlooking the complexities of optical spectra. In this work, we employ deep graph neural networks (GNNs) to predict the frequency-dependent complex-valued dielectric function across the infrared, visible, and ultraviolet spectra directly from crystal structures. We explore multiple architectures for multi-output spectral representation of the dielectric function and utilize various multi-fidelity learning strategies, such as transfer learning and fidelity embedding, to address the challenges associated with the scarcity of high-fidelity DFT data. Additionally, we model key solar cell absorption efficiency metrics, demonstrating that learning these parameters is enhanced when integrated through a learning bias within the learning of the frequency-dependent absorption coefficient. This study demonstrates that leveraging multi-output and multi-fidelity ML techniques enables accurate predictions of optical spectra from crystal structures, providing a versatile tool for rapidly screening materials for optoelectronics, optical sensing, and solar energy applications across an extensive frequency spectrum.
\end{abstract}

\maketitle


\section{Introduction}
Frequency-dependent optoelectronic properties provide essential insights critical for the design and optimization of a wide array of devices spanning various applications including photovoltaic (PV) cells \cite{day2019improving}, light-emitting diodes \cite{ren2021emerging}, transparent electronics \cite{won2023transparent}, optical sensors \cite{li2015plasmon}, optical coatings \cite{hu2019optical}, chemical analysis \cite{babbe2020optical}, and astrochemistry \cite{materese2020laboratory}. The capability to accurately and efficiently predict optical properties across a spectrum of frequencies is critical for integrating materials into cutting-edge optoelectronic devices. Computational approaches, mainly using DFT, can provide optical spectra with accuracy comparable to experiments more cost-effectively. Additionally, DFT optical spectra, generated with consistent calculation settings, can serve as benchmarks to identify influences beyond band-to-band transitions, such as experimental setups or substrate effects. However, the vast array of candidate materials poses formidable computational challenges for DFT, necessitating the exploration of data-driven predictive models for preliminary screening.

Nevertheless, a gap persists in the literature concerning ML surrogate models capable of accurately predicting the frequency-dependent optical properties of solid materials. Previous studies have exclusively concentrated on predicting individual scalar properties, such as the band gap \cite{wang2022accurate, rajan2018machine, zhuo2018predicting} and the static dielectric constant \cite{shimano2023machine, takahashi2020machine}, without accounting for the frequency dependence of optical properties. While the prediction of spectral properties has only recently emerged in materials science, multiple studies have explored multi-output learning for predicting the electronic and phononic density of states \cite{mahmoud2020learning, fung2022physically, kong2022density}. In the context of optical spectra, a hierarchical-correlation model was utilized to predict the absorption coefficient at different frequencies within the visible range, solely based on the chemical composition within a collection of $69$ three-cation metal oxides \cite{kong2021materials}. Another study utilized a Gaussian process model to predict the dielectric constant of polymers using a dataset of $1210$ experimentally-measured values at different frequencies. \cite{chen2020frequency, chen2020frequency}. These studies, however, were confined to specific material chemical spaces, employing composition-based features while excluding crystal structure, and constrained their predictions to particular discrete frequencies. To our knowledge, no published work has yet utilized ML to directly predict the continuous, frequency-dependent dielectric function or absorption coefficient across a general chemical space of materials based on the crystal structure. 

Early endeavors in predicting material properties relied on manual feature engineering from composition, crystal structure, and electronic band structure using featurization algorithms \cite{ward2018matminer}. Currently, state-of-the-art predictive modeling of materials utilizes GNNs \cite{chen2019graph, choudhary2021atomistic, xie2018crystal}, which adeptly generate latent feature representations from composition and structure, enabling automatic learning of features specific to the target property. In this work, we use GNNs to predict the frequency-dependent, complex-valued dielectric function of solid materials directly from crystal structure data. The dielectric function, a fundamental spectral output from ab initio calculations, determines material's response to electromagnetic waves. It also enables the calculation of crucial practical frequency-dependent optical properties, such as the refractive index, electron energy loss spectra \cite{lewis2022forecasting}, quality factors for localized surface plasmon resonances and surface plasmon polaritons \cite{shapera2022discovery}, and the quantum efficiency of optical sensors and PV cells \cite{chander2015study}.

For material spectral properties like phonon or electronic density of states, the full-energy density of occupied states is characterized by a known integral for each material, attributed to its atom or electron count. This facilitates modeling the spectrum as a probability distribution, simplifying learning by correlating increases in intensity in one range with decreases in another. However, the optical spectrum lacks this property, and the optical response can vary significantly in magnitude among materials. Yet, scaling the optical spectrum can still offer a way to establish a correlation within the predicted output. We find that proper spectrum scaling can lead to improved organization within the latent feature space, subsequently enhancing model's performance. 

Furthermore, while training ML models on high-fidelity data yields more accurate results, securing a sufficient volume of such data for effective training presents a notable challenge. A viable solution is adopting multi-fidelity learning frameworks that integrate data from both low-fidelity and high-fidelity sources. The larger datasets employing low-fidelity DFT functionals can enhance GNN models' ability to learn better encodings of crystal structures, consequently boosting performance in learning high-fidelity data. In this study, we investigate two multi-fidelity learning approaches: "transfer learning" and "fidelity embedding" \cite{jha2019enhancing, hoffmann2023transfer, chen2021learning}, demonstrating that multi-fidelity learning effectively addresses the bottleneck caused by the scarcity of high-fidelity optical spectra. 

Finally, we assess the potential to enhance the prediction accuracy of physical features of interest alongside the overall spectrum prediction by incorporating a physical learning bias during training. Focusing on PV cell absorption metrics—short-circuit current, reverse saturation current, and the spectroscopic limit of maximum efficiency (SLME), which represents the theoretical maximum photoconversion efficiency of a single p-n junction PV cell—we demonstrate that learning these properties within the context of frequency-dependent absorption coefficient learning through learning biases is more effective than directly learning them as standalone target properties.

\section{Methods}

\textbf{Data Set:} The dataset was obtained from the JARVIS-DFT Vienna Ab initio Simulation Package (VASP) raw files on Figshare as of January 2024, encompassing all publicly available optical spectra data \cite{wines2023recent, choudhary2020joint, choudhary2018computational}. It comprised $34,327$ calculations employing the OptB88vdW (OPT) functional and $14,560$ calculations utilizing the meta-GGA modified Becke-Johnson (MBJ) potential \cite{klimevs2009chemical, becke2006simple, tran2009accurate}, following data cleansing. The complex dielectric tensor, $\varepsilon(E) = \varepsilon_1(E) + i \varepsilon_2(E)$, is provided for each material in the dataset across a $5,000$-point energy grid that spans the entire range between the minimum and maximum Kohn-Sham eigenvalues. To ensure uniformity in the energy values at which $\varepsilon(E)$ is evaluated for all materials within the dataset, we employ cubic interpolation and then uniformly extract the function values within the $0$ eV to $12.0$ eV range with a resolution of $0.04$ eV. This energy range encompasses the infrared (IR), visible, and ultraviolet (UV) spectral regions. Notably, the employed model frameworks can be readily adapted to cover broader or different spectral ranges. 

\begin{figure*} 
    \centering
    \includegraphics[width=0.96\textwidth]{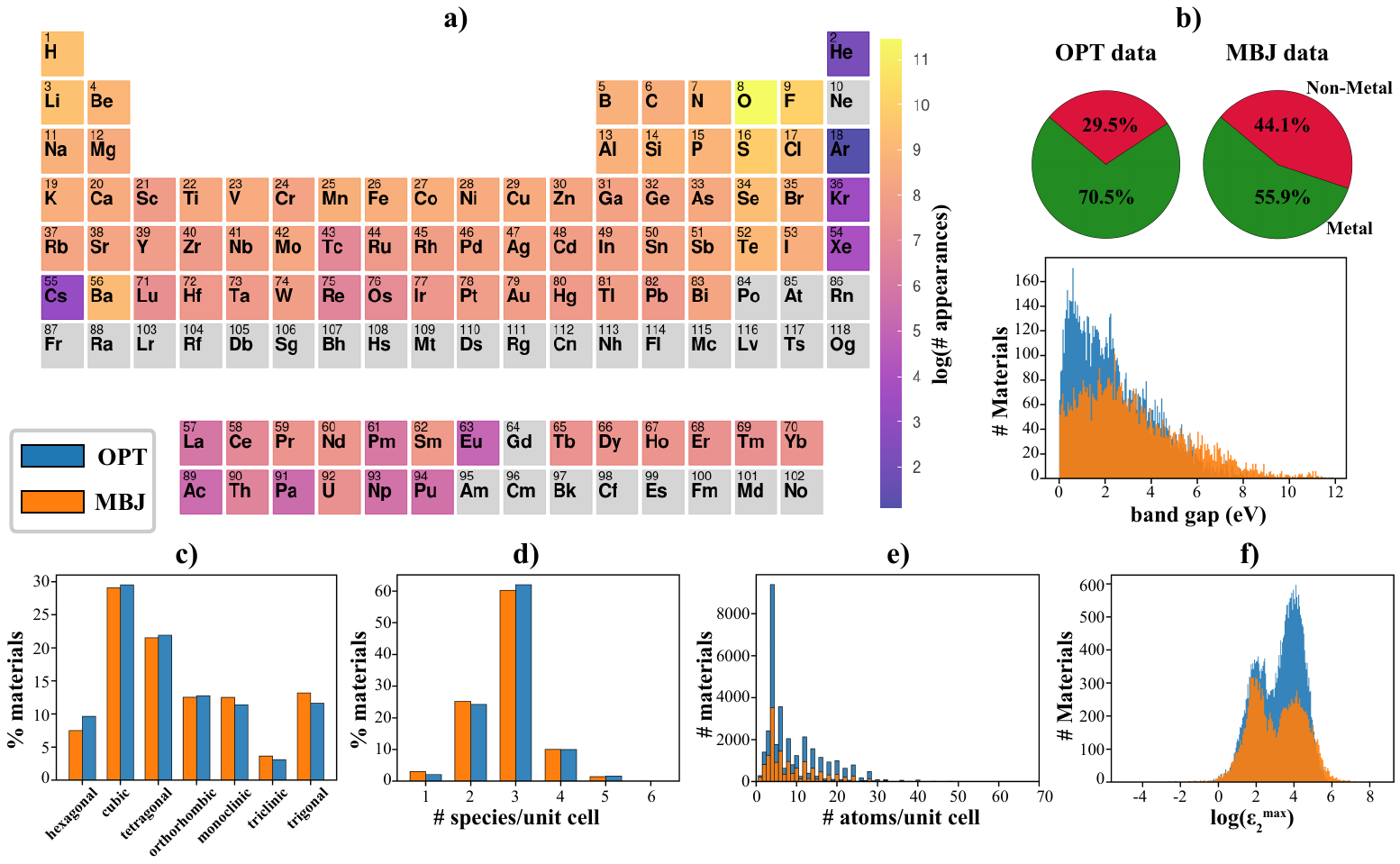}
    \caption{\textbf{Summary of the frequency-dependent optical dataset:} (a) Heatmap illustrating the presence of the periodic table elements in the combined crystal structures of the OPT and MBJ datasets. Elements not present in the dataset are marked in gray. (b) Proportions of metals versus non-metals and distribution of band gap values for non-metals in both datasets. (c-e) Distributions of crystal systems, number of distinct species in the unit cell, and atom count in the unit cell. (f) Histogram distribution of the logarithm of the imaginary dielectric function peak value.}
     \label{fig:fig1}
\end{figure*}

The $\varepsilon(E)$ output from VASP is represented by a $3 \times 3$ tensor, computed for the primitive unit cell of each material. To simplify the problem, we diagonalize $\varepsilon(E)$, yielding three eigenvalues, which correspond to eigenvectors oriented along the principal crystallographic axes. Our GNN models are then trained using the mean of these eigenvalues. It is important to note that predicting the optical spectra along a specific symmetry axis follows the same formalism. For example, in the context of van der Waals layered materials, the model can be trained using either the eigenvalues of the in-plane or out-of-plane axis. To streamline learning, we utilize the imaginary part of the dielectric function ($\varepsilon_2(E)$) and a reduced form of the absorption coefficient, rather than the real part ($\varepsilon_1(E)$), to model the complex optical spectra. This choice is made due to commonalities shared by the imaginary part and the absorption coefficient, such as their non-negativity and zero values for energies within the band gap. The reduced adsorption coefficient is defined as 
\begin{equation}
\alpha(E) = -\varepsilon_1(E)+\sqrt{\varepsilon_1^2(E)+\varepsilon_2^2(E)}
\end{equation}
The total absorption coefficient is given by $2 \sqrt{2} \pi (E/hc) \sqrt{\alpha (E)}$, where $h$ and $c$ represent Planck's constant and the speed of light in vacuum, respectively. Notably, the real part can be readily derived from both the imaginary part and the absorption coefficient. Figure \ref{fig:fig1} illustrates a statistical distribution of all materials within the extracted OPT and MBJ datasets, categorized according to the frequency of elements they contain, band gap values, crystal systems, diversity of constituent species, number of atoms in the unit cell, and $\varepsilon_2$ peak magnitudes. More details about the dataset preparation are provided in the Supporting Information (SI).

\begin{figure*} 
    \includegraphics[width=0.96\textwidth]{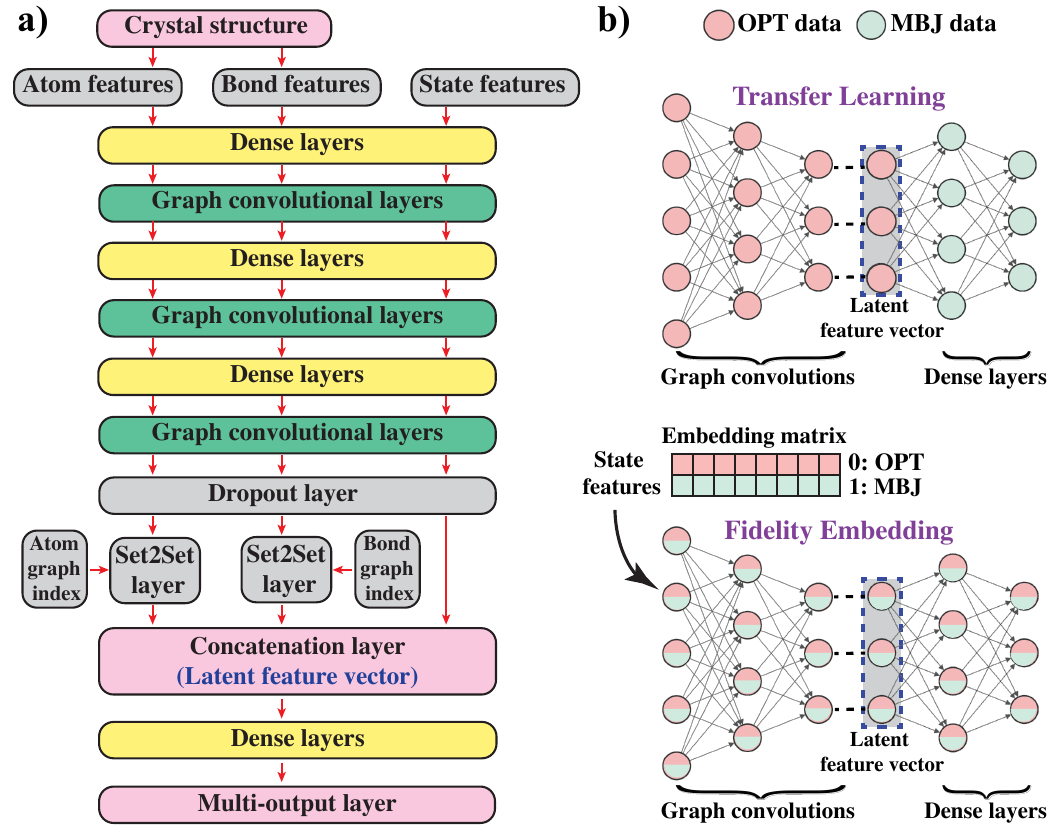}
    \caption{\textbf{Overview of the GNN architecture and multi-fidelity learning frameworks:} a) Schematic of the GNN architecture, illustrating input structures with embedded atomic numbers, Gaussian-expanded bond distances, and state features. The multi-output layer outputs the predicted discretized optical spectrum over the considered frequency range. b) Diagram of the multi-fidelity learning frameworks: Transfer learning sequentially trains on OPT data across the entire GNN, then employs the learned LFV as input to the dense layers for subsequent MBJ data learning, while freezing the graph convolutional layers; fidelity-embedding jointly learns OPT and MBJ data throughout the entire GNN, using a trainable embedding matrix as the input state feature to encode the DFT functional fidelity level (MBJ or OPT) for each crystal structure.}
     \label{fig:fig2}
\end{figure*}

\textbf{Graph Neural Network Formalism:} Figure \ref{fig:fig2}-a) illustrates a general architectural overview of the employed GNN models. We utilize the MEGNet graph convolutional layers proposed by C. Chen \textit{et al.} \cite{chen2019graph}. Initially, each crystal structure is transformed into a graph characterized by node features, edge features, and global features, corresponding to atoms, bonds, and the overall state of the structure, respectively. For atom (node) features, we use only the atomic numbers of the constituent elements, which are then mapped to an embedding layer to learn elemental embeddings. Bond (edge) features are expressed through expanding interatomic bond distances using a Gaussian basis with 180 centers, each $0.5$ \AA{} wide, uniformly distributed between bonded atoms. Atoms are considered bonded if their interatomic distance is less than a cutoff radius of $5.5$ \AA{}, which encompasses not only the nearest neighbors but also interactions with more distant atoms. For state (global) features in fidelity-embedding GNN models, which are jointly trained on multi-fidelity data, the fidelity level of the DFT functional (OPT or MBJ) associated with each data point is represented by an integer (0 or 1). This is followed by a trainable embedding layer utilized to learn encodings for the fidelity levels. In contrast, for single-fidelity (SF) learning—where the model is trained on data from a single DFT functional—or in transfer learning—where the model is trained on multi-fidelity data sequentially—only two placeholder nodes, without embedding, are employed to facilitate global information exchange.

The input features undergo a preprocessing step through dense layers before being forwarded to the consecutive graph convolutional layers. These convolutional layers execute a sequence of update operations through convolution and pooling layers, transforming an input graph $G = (\mathbf{e}, \mathbf{v}, \mathbf{u})$ into an output graph $G' = (\mathbf{e'}, \mathbf{v'}, \mathbf{u'})$, where $\mathbf{e}$, $\mathbf{v}$, and $\mathbf{u}$ represent the edge, vertex, and global features, respectively. A stack, comprising three repetitions of the preprocessing dense layers and graph convolutional layers as shown in Figure \ref{fig:fig2}-a), is utilized to enhance model flexibility and indirectly access information beyond the $5.5$ \AA{} cutoff radius, enabling the model to capture more intricate long-range interactions \cite{chen2019graph}. A dropout layer follows the final graph convolution layer to mitigate overfitting. Rather than padding the structure graphs to uniformize the sizes of their atomic features, they are assembled into a single, large disjoint union graph for training. In the final stages of the model, a readout operation is performed on both the atom and bond feature vectors using an order-invariant set2set model \cite{chen2019graph, vinyals2015order}. The set2set layer combines atom and bond feature vectors with vectors denoting the indices of these atoms and bonds within the disjoint union graph. After the readout process, the atom, bond, and state feature vectors are concatenated to form the latent feature vector (LFV), which is subsequently processed through a series of dense layers to generate the multi-output prediction representing the discretized optical spectrum over the considered frequency range. Further details concerning the GNN architecture, hyperparameters, and MEGNet graph convolutional layers are available in the SI.

\textbf{Model Training:} The model construction and training was executed though employing the Keras API with the TensorFlow backend \cite{tensorflow2015-whitepaper, chollet2015keras}. Our dataset was partitioned into three segments, with $80\%$ allocated for training, $5\%$ for validation, and $15\%$ for testing. The selection of hyperparameters, including the size of atom and bond features, dimensions of hidden layers, batch size, and dropout rate was meticulously chosen through Bayesian hyperparameter optimization facilitated by Optuna \cite{akiba2019optuna} (See the SI). The best-performing models were selected based on their performance on the validation set and subsequently evaluated using the test set. 

Training on multi-fidelity data can be approached through various methods. Figure \ref{fig:fig2}-b) presents a schematic of the two multi-fidelity frameworks considered in this study: (1) transfer learning (TL) and (2) fidelity embedding (FE). In the TL framework, multi-fidelity learning progresses sequentially: the GNN model is initially trained on all of the OPT data, and then the LFV is obtained. Following this, the set of dense layers after the LFV is further optimized to accommodate additional layers and neurons, and then trained on the MBJ dataset using an $80/5/15$ train/validation/test split. This means that the weights of the layers before the LFV remain frozen, and only the later dense layers of the GNN are trained on the high-fidelity MBJ data. In the FE framework, multi-fidelity learning is approached jointly. The fidelity level (MBJ or OPT) for each data point is encoded as an integer and inputted into the GNN model through a trainable fidelity-embedding matrix, serving as the input state feature vector. Optimal validation results for $\varepsilon_2$ and $\alpha$ were achieved with fidelity-embedding vector lengths of $20$ and $16$, respectively. The $80/5/15$ train/validation/test split applies exclusively to the MBJ dataset, while the entire OPT dataset is used for training. Further details regarding the architecture of the TL and FE GNNs are available in the SI. \\

\section{RESULTS AND DISCUSSION}

\begin{figure*} 
    \centering
    \includegraphics[width=0.96\textwidth]{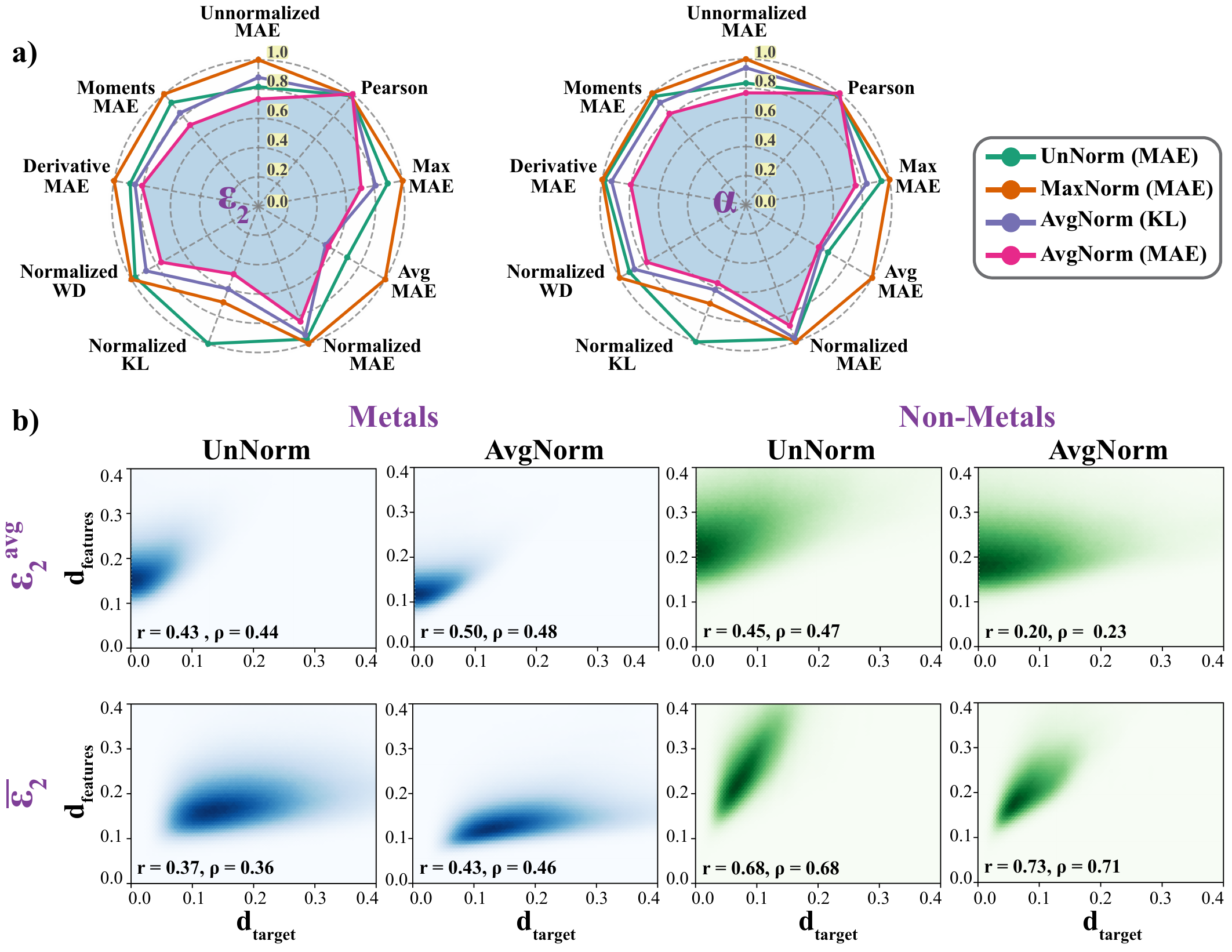}
    \caption{\textbf{Effect of the multi-output architecture on model performance:} a) Radar charts illustrating the median errors, normalized to a maximum of one, for the four considered multi-output GNN architectures. These GNN models differ in their spectrum scaling method and/or the loss function. The shown performance evaluations are conducted using the MBJ data test, detailing various metrics for both the imaginary part ($\varepsilon_2$) and the reduced absorption coefficient ($\alpha$). Numeric values of the median error and the IQR are outlined in Tables S1 and S2. b) The Pearson ($r$) and Spearman ($\rho$) correlation coefficients between the normalized Euclidean pairwise distances of $\varepsilon_2$ targets (which include average norms ($\varepsilon_2^{\text{avg}}$) and PDF-normalized spectra ($\overline{\varepsilon}_2$)) and those of latent feature vectors are depicted for both metals and non-metals in the whole MBJ dataset.}
     \label{fig:fig3}
\end{figure*}

\textbf{Spectrum Multi-Output Architecture:} In this section, we evaluate the performance of different multi-output GNN architectures for representing the optical spectrum, which involve various combinations of data scaling schemes and training loss functions. All GNN models discussed in this section are solely trained and evaluated utilizing MBJ data. Since we have no ground truth regarding a physical scaling feature, such as electron/atom number for the electronic/phonon density of states, we explore both scaling to a maximum value of $1$ (MaxNorm) and normalization to a cumulative sum of $1$ (AvgNorm), alongside the unnormalized spectrum (UnNorm) \cite{kong2022density, fung2022physically}. For the MaxNorm and AvgNorm models, the loss function is formulated as $\mathcal{L} = \mathcal{L}_N + w \mathcal{L}_{S}$, where $\mathcal{L}_N$ pertains to the error in the norm, $\mathcal{L}_{S}$ to the error in the normalized spectrum, and $w$ is a hyperparameter that denotes the relative weight of the two loss components during training, and its optimal value is determined through a grid search process on the validation set (See SI). The mean absolute error (MAE) loss function is utilized for the UnNorm model, for $\mathcal{L}_N$ in both MaxNorm and AvgNorm models, and for $\mathcal{L}_{S}$ in the MaxNorm model. In the AvgNorm model, where the normalized spectrum represents a probability density function (PDF), we experiment with training $\mathcal{L}_{S}$ using two distinct loss functions: MAE and the Kullback-Leibler (KL) divergence loss \cite{Joyce2011}. The four models \{UnNorm, MaxNorm, AvgNorm (KL), AvgNorm (MAE)\} share the same architecture, differing only in the output layer and/or the loss function. In the UnNorm model, the output layer comprises a dense layer of $300$ neurons (representing the considered $12$ eV range at $0.04$ eV resolution) and features a Rectified Linear Unit (ReLU) activation. For the remaining three models, the output layer consists of $301$ neurons, including an additional neuron representing the norm with ReLU activation. For MaxNorm, a sigmoid activation function is employed for the normalized spectrum ($S$), specifically applied to ($S$-$S_{\text{max}}/2$). For the two AvgNorm models, the normalized spectrum utilizes a softmax activation function applied to ($S$-$S_{\text{max}}$). The radar plots in Figure \ref{fig:fig3}-a) summarize the performance of the four models on the $\varepsilon_2$ and $\alpha$ MBJ test sets. The benchmarking metrics cover various indicators to evaluate performance across different facets of the spectrum. These metrics include, in order: the overall MAE and Pearson correlation for the unnormalized spectrum ($\varepsilon_2$ and $\alpha$), MAE for the maximum and average values, along with MAE, KL divergence, Wasserstein distance (WD) \cite{frogner2015learning}, and first derivative MAE for the PDF-normalized spectrum ($\overline{\varepsilon}_2$ and $\overline{\alpha}$). Additionally, we define another metric for aggregated statistical moments, for instance, for $\overline{\varepsilon}_2$, as $\sum_{n} (E^n \cdot \overline{\varepsilon}_2)^{(1/n)}$, encompassing the first four moments. Numeric values of the radar plots, including uncertainties represented by the interquartile range (IQR), are detailed in Tables S1 and S2.

Figure \ref{fig:fig3}-a) shows that the AvgNorm (MAE) model consistently outperforms the other three models across all metrics. This superiority can be attributed to several factors. First, the model benefits from extracting the average norm and the transformation of the optical spectrum into a PDF. Figure \ref{fig:fig1}-f) indicates that the peak values in $\varepsilon_2$ span approximately six orders of magnitude in our data set. Thus, this scaling approach enables the model to effectively capture trends in the normalized spectra, thereby preventing materials with high average values from disproportionately influencing the training process. Second, the utilization of the softmax activation function boosts the model's performance by facilitating improved learning of the spectral distribution by constraining the sum of predicted values of the normalized spectrum to unity. This constraint introduces a correlation along the predicted spectrum, where a high probability in some range of the spectrum automatically leads to a reduction in the probability in other ranges. Furthermore, the exponential function in softmax activation responds to lower stimulations (present in the lower regions of the normalized spectrum) with a more uniform distribution, while exponentially amplifying higher stimulations, such as peaks and near-peak regions of the normalized spectrum. This mechanism ensures that large probabilities predominate, while still incorporating information from lower probabilities within the spectrum.

Training the PDF using KL divergence loss is observed to yield similar performance, albeit slightly lower than when employing MAE loss, which, surprisingly, results in a lower KL error on the test set compared to training directly with KL divergence loss. This can be attributed to KL divergence quantifying discrepancies between PDFs logarithmically, helping to enhance sensitivity to low-probability areas but simultaneously reducing it in high-probability regions due to the logarithmic function's dampening effect on large values. Thus, KL divergence may overlook finer details around peak areas within the optical PDF. 

We also experimented with the WD loss for the PDF-normalized spectrum within the AvgNorm architecture but observed that the predicted spectra became impractically noisy. This likely stems from WD penalizing errors in the cumulative distribution function (CDF) rather than the PDF \cite{ramdas2017wasserstein}, leading to precise resonant peak positioning but significant noise in regions such as the baseline (ranges of no interaction with electromagnetic fields) or peak tails. This occurs because the CDF loss permits fluctuations between underestimation and overestimation at successive spectral points, resulting in a lower CDF error comparable to a prediction that consistently under- or overestimates. Given the necessity for accurate, noise-free spectral predictions for analyzing the dielectric function, WD was deemed unsuitable as a standalone loss function for the normalized spectrum and was therefore excluded from our analysis. 

On the other hand, extracting the maximum value and normalizing the spectrum to a maximum of one results in inferior performance compared to the unnormalized case. This decline can partly be attributed to the use of the sigmoid function with the MaxNorm model. Although the sigmoid function offers a smooth exponential form that confines the normalized spectrum within the range of $0$ to $1$, unlike softmax, it does not impose constraints that can correlate the spectrum points. The sigmoid activation for $(S - S_{\text{max}}/2)$ treats values around the half-maximum almost linearly but rapidly saturates values deviating upward from the half-maximum to $1$ and values deviating downward from the half-maximum to zero.  This saturation effect reduces the resolution between points of higher values in the spectrum, thereby contributing to the observed decrease in performance. We also conducted experiments where the application of the sigmoid function in the MaxNorm architecture was omitted, opting instead for a linear activation function with clamping between $0$ and $1$. However, the resulting output did not yield spectra that were deemed to be plausibly smooth. Further discussion providing more insights into the reasons behind the observed effects of spectrum scaling on GNN performance is discussed in the "Latent Space Visualization" section.

The efficacy of the AvgNorm model architecture implies that representing the optical spectrum of arbitrary crystal structures through an average norm, alongside a PDF, learned via uniform error penalization without weighting any part of the distribution more heavily than another while using softmax activation, enhances the model's capacity to discern fundamental patterns in optical spectra across diverse materials. Consequently, the AvgNorm (MAE) architecture is utilized in subsequent analyses. Notably, the ramifications of this scaling extend beyond mere post-processing of the output, profoundly influencing the arrangement of materials in the latent feature space. This phenomenon is illustrated in Figure \ref{fig:fig3}-b), where we detail the distribution of normalized Euclidean pairwise distances for $\varepsilon_2$ targets, including average norms ($\varepsilon_2^{\text{avg}}$) and PDF-normalized spectra ($\overline{\varepsilon}_2$), versus those of the LFVs across all materials in the MBJ dataset, correlating them using both Pearson ($r$) and Spearman ($\rho$) correlation coefficients. In the case of metals, the AvgNorm model is observed to increase $r$ between features and both $\varepsilon_2^{\text{avg}}$ and $\overline{\varepsilon}_2$ by about $16\; \%$. Conversely, for non-metals, while the AvgNorm model increases $r$ between features and $\overline{\varepsilon}_2$ by around $7\; \%$, it concurrently significantly reduces the correlation with $\varepsilon_2^{\text{avg}}$ by about $56\; \%$. These alterations in the correlation of features with targets underscore the profound influence of optical spectrum scaling on the structural organization of the latent feature space. In essence, the AvgNorm model orchestrates a rearrangement of the materials within the latent space, fostering proximity among materials exhibiting similar PDF-normalized spectra, and simultaneously boosting/attenuating the arrangement based on the average norms for metals/non-metals. Further discussion on the organization of materials in the latent space is presented in the "Latent Space Visualization" section.

\textbf{Learning from Multi-Fidelity Data:} Ideally, ML model training should rely on experimental or high-fidelity computational data. However, given the inherent scarcity of such high-fidelity data, practical limitations necessitate leveraging the ample low-fidelity data to enhance predictive power through improved feature learning. 

\begin{figure*} 
    \centering
    \includegraphics[width=0.96\textwidth]{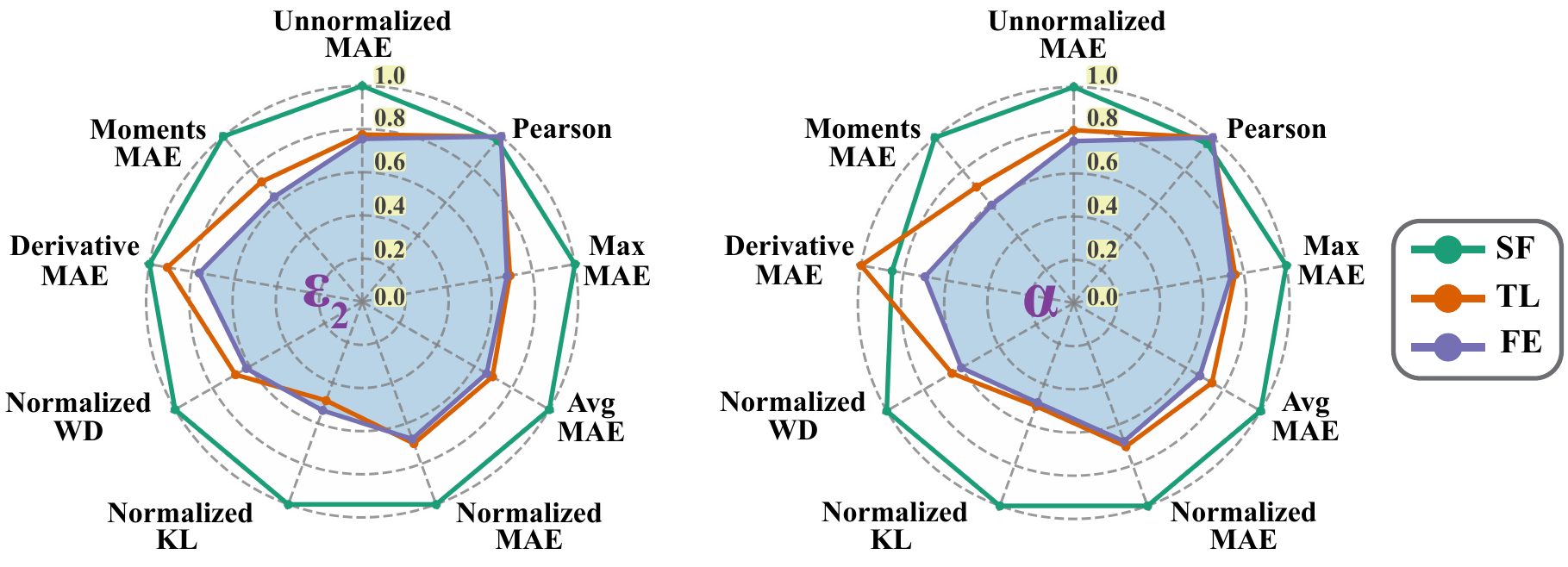}
    \caption{\textbf{Multi-fidelity learning:} Radar charts depicting the median errors, normalized to a maximum of one, for three GNNs: the single-fidelity (SF) model (trained exclusively on MBJ data), the transfer learning (TL) model (initially trained on all OPT data, followed by further training of the dense layers post-LFV on $80\%$ of MBJ data), and the fidelity-embedding (FE) model (trained jointly on all OPT data and $80\%$ of MBJ data). The shown performance evaluations are conducted using the higher-fidelity MBJ data test set, detailing various metrics for both the imaginary part ($\varepsilon_2$) and the reduced absorption coefficient ($\alpha$). Numeric values of the median error and the IQR are outlined in Tables S3 and S4.}
     \label{fig:fig4}
\end{figure*}

The radar plots in Figure \ref{fig:fig4} illustrate the performance metrics for the SF GNN, which is trained solely on MBJ data, as well as for the TL and FE GNNs for both $\varepsilon_2$ and $\alpha$ (numeric values of median error and IQR are detailed in Tables S3 and S4). All three models employ the AvgNorm (MAE) architecture discussed in the previous section. The results indicate a noticeable decrease across all error metrics and an increase in correlation with the DFT (MBJ) spectra upon the incorporation of the lower-fidelity OPT data. Notably, the FE model demonstrates a higher improvement, with the median MAE of the unnormalized spectrum decreasing by $24.6\;\%$ and $25.0\;\%$ for $\varepsilon_2$ and $\alpha$, respectively, compared to reductions of $22.5\;\%$ and $20.0\;\%$ for TL. Both FE and TL exhibit comparable improvements in the Pearson correlation, with $\varepsilon_2$ increases of $2.5\;\%$ for FE versus $2.3\;\%$ for TL, and $\alpha$ increases of $3.9\;\%$ for FE versus $3.7\;\%$ for TL. The superior performance of the FE framework can be attributed to its joint learning strategy, which simultaneously integrates low- and high-fidelity data during training, in contrast to the sequential learning approach utilized in TL. In TL, only the dense layers post-LFV can detect the nuanced differences between OPT and MBJ data. Conversely, FE empowers the entire GNN to fine-tune its weights for fitting both OPT and MBJ data, thus achieving a broader optimization scope. Moreover, while the larger size of the OPT dataset enables the TL model to glean a more robust LFV compared to SF models trained solely on the smaller-sized MBJ data, this approach inherently restricts the LFV learning to patterns of the lower-fidelity dataset, potentially overlooking insights that could be gained from holistic learning involving both data fidelities simultaneously. Therefore, the joint learning strategy of the FE framework enables the GNN to exploit the most extensive dataset resulting from the amalgamation of OPT and MBJ datasets, thereby accessing a broader range of information and achieving a more refined LFV compared to TL.

\textbf{Latent Space Visualization:} In essence, the enhanced performance of a GNN model in predicting material properties implies an improved ordering of materials within the latent feature space. While this concept is relatively straightforward in single-target prediction scenarios, where an improved latent space representation should manifest a more correlated ordering of latent feature vectors with the target scalar, complexities emerge in spectral multi-output prediction problems. In such cases, numerous scalar and vectorial features derived from the spectrum can potentially organize the latent space. The pertinent question then becomes: which feature holds greater significance in organizing the latent space? 

\begin{figure*} 
    \centering
    \includegraphics[width=0.96\textwidth]{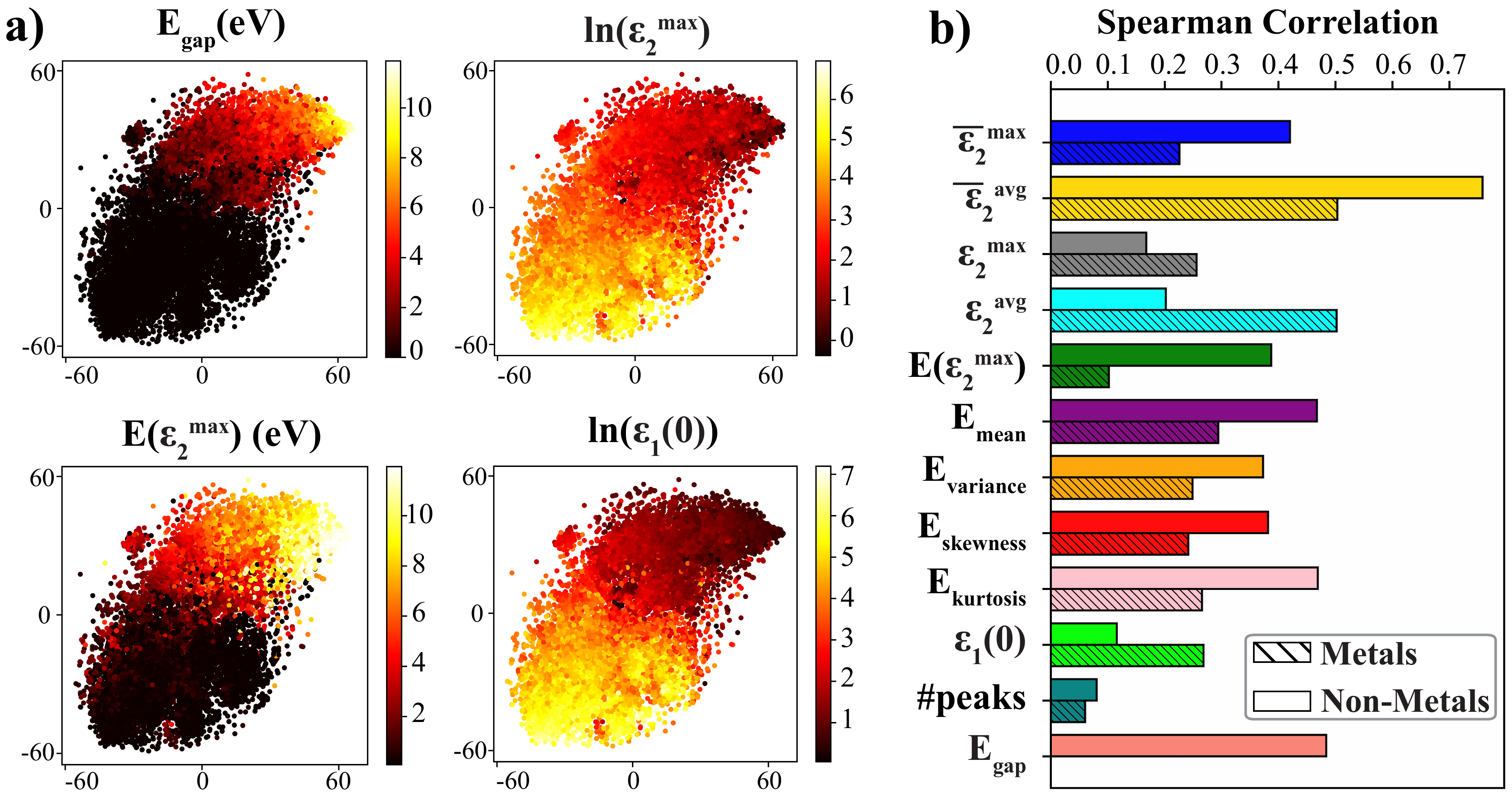}
    \caption{\textbf{Significance of optical spectrum properties in organizing the latent feature space of crystal structures:} a) Two-dimensional t-SNE projection of the latent features of all the structures in the MBJ dataset (perplexity = 150). The color maps represent DFT (MBJ) values for various scalar properties of the multi-output optical spectrum. The latent feature vectors are derived from the AvgNorm FE model optimized for predicting $\varepsilon_2$. b) Spearman correlation coefficients between the normalized Euclidean pairwise distances of latent feature vectors and those for different properties extracted from the optical spectrum for both metals and non-metals.} 
     \label{fig:fig5}
\end{figure*}

To gain deeper insights into the latent feature space, Figure \ref{fig:fig5}-a) illustrates a two-dimensional projection via t-distributed stochastic neighbor embedding (t-SNE) \cite{JMLR:v9:vandermaaten08a} of the latent features of all structures within the MBJ dataset, accompanied by heatmaps for certain physical scalar properties derived from the optical spectrum. The latent feature vectors are derived from the best-performing AvgNorm FE GNN model optimized for predicting $\varepsilon_2$. Further quantitative insights into the latent feature space are obtained by calculating the Spearman rank-order correlation between the normalized Euclidean pairwise distances of latent features and the corresponding normalized Euclidean pairwise distances of various target properties extracted from the optical spectrum (0-12 eV range), as illustrated in Figure \ref{fig:fig5}-b).

The band gap (E$_{\text{gap}}$) heatmap t-SNE plot in Figure \ref{fig:fig5}-a) reveals that the GNN model proficiently segregates metals from non-metals into two distinctly discernible clusters within the latent space, thereby showcasing the model's adept comprehension of the distinctive optical characteristics of these two material categories. Moreover, a prominent gradient in E$_{\text{gap}}$ is evident within the non-metal cluster, capturing the diversity in band gaps among semiconductors and insulators. This is demonstrated by a Spearman correlation of $\rho = 0.48$ between E$_{\text{gap}}$ and the latent features, as illustrated in Figure \ref{fig:fig5}-b). In a similar vein, the model exhibits structured organization within the latent space regarding additional dielectric properties, such as the logarithm of the peak value of the imaginary dielectric function, denoted by ln($\varepsilon_2^{\text{max}}$), and the corresponding energy at this peak, denoted as E($\varepsilon_2^{\text{max}}$). A pronounced gradient in ln($\varepsilon_2^{\text{max}}$) is evident within the metal cluster, characterized by markedly elevated peaks at its boundary, distant from the interface between metal and non-metal clusters. This pattern reveals that metals distant from the metal/non-metal cluster interface display elevated optical conductivity peaks, which diminish progressively toward the interface, running counter to the band gap gradient direction within the non-metal cluster; this is consistent with physical expectations, as one would expect optical conductivity to exhibit patterns that are opposite to those of the band gap. Conversely, within the non-metal cluster, a less pronounced gradient is observed for ln($\varepsilon_2^{\text{max}}$). This can be attributed to the fact that the imaginary dielectric function of semiconductors and insulators exhibits a more complex dependence on band structure, reflecting the intricate probabilities of valence to conduction interband transitions. In contrast, the simpler intraband transitions of free electrons in metals, particularly at lower energies, can be effectively characterized using bulk properties such as electron density and scattering rate, as described by the Drude model \cite{hilfiker2018dielectric}. Therefore, the optical spectra of non-metals are more challenging to represent with simple dielectric magnitudes. This is highlighted by the lower correlation coefficients of both maximum and average values ($\varepsilon_2^{\text{max}}$ and $\varepsilon_2^{\text{avg}}$) for non-metals compared to metals, as demonstrated in Figure \ref{fig:fig5}-b), where non-metals exhibit correlations of $0.17$ and $0.20$, compared to $0.26$ and $0.50$ for metals. 

Furthermore, a gradient is evident in E($\varepsilon_2^{\text{max}}$), with metals generally manifesting lower values as indicated in Figure \ref{fig:fig5}-a), indicative of their lower natural frequencies due to the presence of delocalized electrons. Conversely, semiconductors and insulators tend to exhibit notably higher E($\varepsilon_2^{\text{max}}$) values owing to the increased energies required to excite their tightly-bound valence electrons to the conduction band, necessitating the overcoming of the band gap before observing the peak. Interestingly, within the metal cluster, the latent space of the imaginary part displays a noticeable gradient in the logarithm of the static dielectric constant, ln($\varepsilon_1(0)$), which is related to the real part. This suggests that the GNN effectively captures the physical linkage between the imaginary and real parts of the dielectric function via the Kramers-Kronig relation \cite{kuzmany1998dielectric}. In contrast, a less pronounced gradient is observed among non-metals, again highlighting their more intricate dependence on band structures rather than simple dielectric magnitudes. This is reflected by a lower $\varepsilon_1(0)$ correlation value of $0.12$ for non-metals compared to $0.27$ for metals, as shown in Figure \ref{fig:fig5}-b). 

To address our question regarding which feature from the optical spectrum is most influential in structuring the latent space, an examination of Figure \ref{fig:fig5}-b) indicates that for non-metals, the average-normalized spectrum ($\overline{\varepsilon}_2^{\text{avg}}$), interpreted as a PDF, emerges as the property most strongly correlated with the latent features, exhibiting a high $\rho$ of $0.76$, compared to the other evaluated properties, none of which surpass $0.48$. Notably, the pronounced correlation of $\overline{\varepsilon}_2^{\text{avg}}$ is not a consequence of training the FE GNN model with the AvgNorm architecture. Evidence supporting this observation is illustrated in Figure S6, which shows $\overline{\varepsilon}_2^{\text{avg}}$ maintaining its status as the most correlated property for non-metals across all SF models, regardless of the employed output scaling architecture. Thus, the PDF-normalized $\overline{\varepsilon}_2^{\text{avg}}$ can be deemed as a fundamental descriptor for learning the optical spectra of semiconductors and insulators, reinforcing our earlier observations regarding the superior performance of the AvgNorm model compared to the UnNorm model. Moreover, the lower correlation of features with the max-normalized spectrum ($\overline{\varepsilon}_2^{\text{max}}$), with $\rho = 0.42$ for non-metals, bolsters the previously observed superior performance of the AvgNorm architecture over the MaxNorm. E$_{\text{gap}}$ and statistical scalar attributes, including E$_{\text{mean}}$, E$_{\text{variance}}$, E$_{\text{skewness}}$, E$_{\text{kurtosis}}$, and E$(\varepsilon_2^{\text{max}})$ (indicating the mode), rank as secondary in significance for structuring non-metals' latent space, exhibiting $\rho$ values between approximately $0.37$ and $0.48$. Nonetheless, these attributes are comprehensively integrated within the normalized spectrum, $\overline{\varepsilon}_2^{\text{avg}}$. Additional scalar properties, including dielectric magnitudes such as the static dielectric constant ($\varepsilon_1(0)$), the average ($\varepsilon_2^{\text{avg}}$) and maximum ($\varepsilon_2^{\text{max}}$) imaginary dielectric values, and other generic features like the number of spectral peaks (defined as the largest local maxima exceeding $0.25$ of the highest peak and separated by at least $2$ eV), exhibit the lowest correlation scores with the latent features, all with $\rho \leq 0.2$. This underscores their relatively minor role in shaping the latent space of non-metals. 

The correlation landscape for the latent space of metals exhibits a distinct profile, with both $\overline{\varepsilon}_2^{\text{avg}}$ and $\varepsilon_2^{\text{avg}}$ playing equally significant roles in structuring the latent space, each exhibiting a correlation value of $\rho = 0.50$. As discussed, the spectral dielectric response of metals, characterized by predominant intraband transitions at low energies within the IR range, allows a simple dielectric magnitude, specifically $\varepsilon_2^{\text{avg}}$, to serve as a key descriptor for metals, emphasizing their optical conductivity. However, similar to non-metals, $\overline{\varepsilon}_2^{\text{avg}}$ remains crucial for defining the latent space and accurately representing the spectral energy distribution. This consideration becomes particularly relevant given that metals may exhibit additional peaks at higher frequencies due to potential interband transitions in the late-visible and UV ranges. Consequently, $\varepsilon_2^{\text{avg}}$ and $\overline{\varepsilon}_2^{\text{avg}}$ are identified as two pivotal descriptors for effectively learning metals' optical spectra.

\begin{figure*} 
    \centering
    \includegraphics[width=0.96\textwidth]{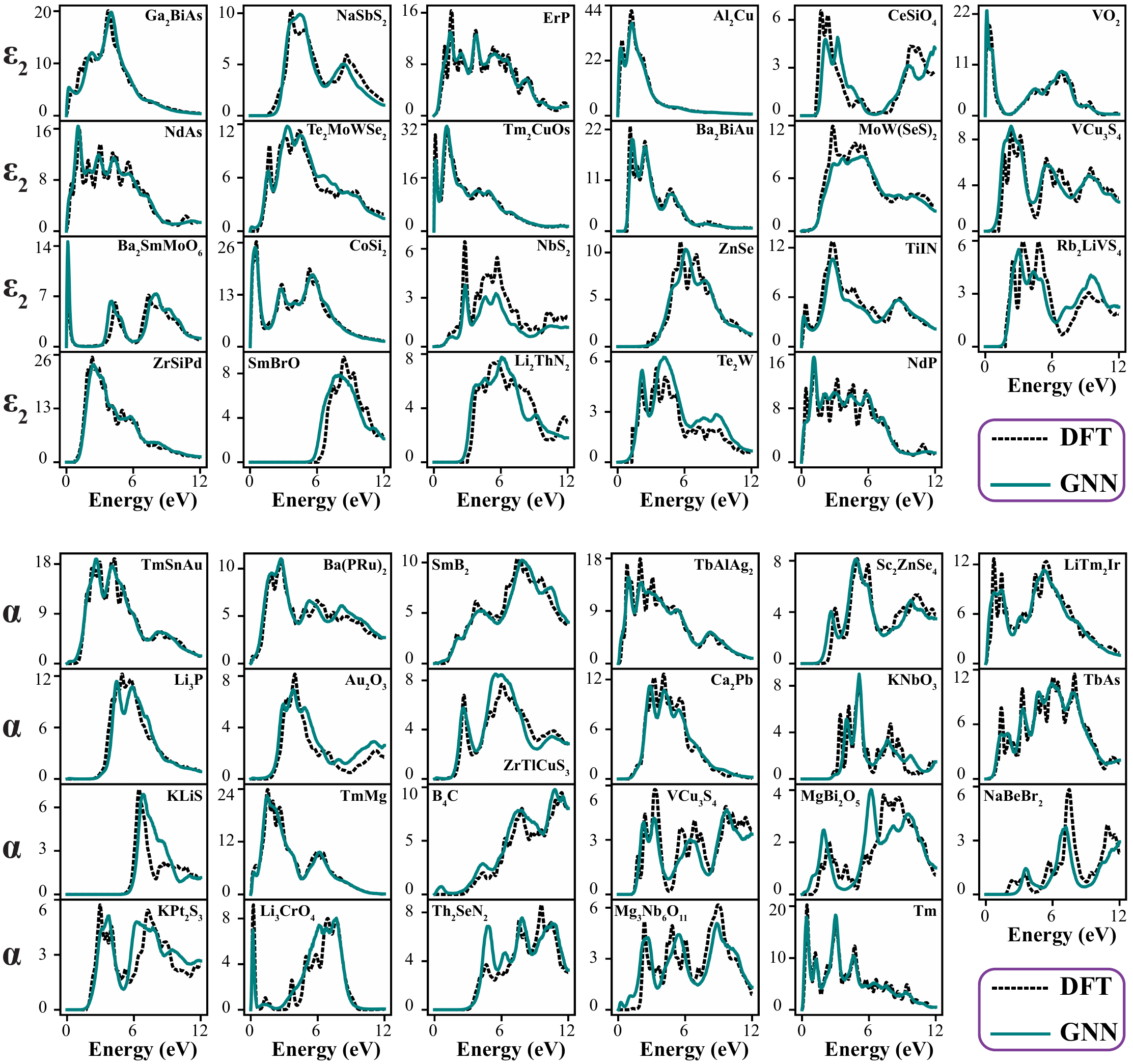}
    \caption{\textbf{Graph neural network predictions of optical spectra:} The interpolated predictions for the dielectric function's imaginary part ($\varepsilon_2$) and the reduced absorption coefficient ($\alpha$) are shown against the DFT interpolations for samples of materials obtained from the MBJ test set within $\pm 5\;\%$ around median MAE in the unnormalized spectrum. The predictions are generated using the optimized AvgNorm (MAE) FE models trained on all OPT data and $80\;\%$ of the MBJ data.}
     \label{fig:fig6}
\end{figure*}

\textbf{Prediction of Frequency-Dependent Optical Spectra \& Solar Absorption Efficiency:} 
Figure \ref{fig:fig6} depicts predictions for both $\varepsilon_2$ and $\alpha$ for materials selected from the MBJ test set, chosen within an error margin of $\pm 5\;\%$ around the median MAE error in the unnormalized spectrum. These predictions are generated using the optimized AvgNorm (MAE) FE GNN models. The shown median performance highlights the efficacy of the GNN models, enhanced by optimized spectrum multi-output scaling and multi-fidelity learning, in accurately capturing the nuances in DFT optical spectra at the meta-GGA MBJ level. Notably, the models exhibit proficiency in precisely identifying peak values and their respective positions across the entire spectrum, including the IR, visible, and UV regions, spanning metals, semiconductors, and insulators. From $\varepsilon_2$ and $\alpha$, $\varepsilon_1$ can be derived, yielding the complex dielectric function, from which various significant frequency-dependent optical properties can be calculated.

Having presented GNN models capable of generalizing across a broad range of frequencies and diverse materials, it is noteworthy that specific practical applications often necessitate focusing on particular material groups within narrower spectral regions. For example, in solar cell applications, the emphasis is on semiconductors with absorbance profiles efficiently capturing incident solar irradiation. The SLME represents the theoretical maximum photoconversion efficiency of a single p-n junction solar cell. While SLME is a scalar property directly learnable through a single-output ML model, gaining insight into the spectral absorbance characteristics contributing to this SLME value is valuable. This deeper understanding can clarify why a material exhibits lower or higher SLME, suggesting potential modifications or exploration paths. Therefore, predicting SLME by forecasting the absorption coefficient proves advantageous.

\begin{figure*} 
    \centering
    \includegraphics[width=0.96\textwidth]{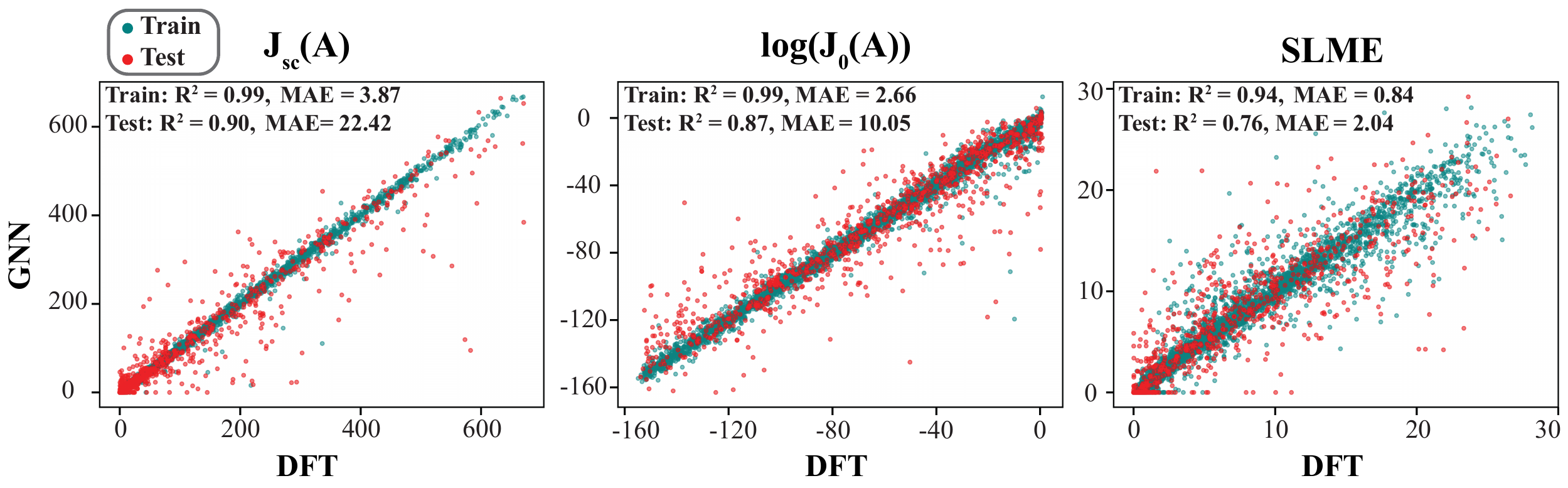}
    \caption{\textbf{Graph neural network prediction of solar energy absorption efficiency:} Predictions of the short-circuit current (J$_\text{sc}$), natural logarithm of the reverse saturation current (log(J$_0$)), and the spectroscopic limit of maximum efficiency (SLME) are validated for materials with band gaps ranging from $0.1$ to $4.5$ eV against DFT (MBJ) values. J$_\text{sc}$ and J$_0$ are expressed in amperes (A). The employed GNN predicts the frequency-dependent absorption coefficient with extra learning bias to emphasize learning J$_\text{sc}$ and log(J$_0$).}
    \label{fig:fig7}
\end{figure*}

We demonstrate that a multi-output GNN model, trained to predict the absorption coefficient, can accurately forecast SLME at levels comparable to, or even surpassing, those of a single-output GNN model specifically designed for SLME learning, while maintaining precise spectrum prediction capabilities. By restricting our OPT and MBJ data to materials with band gaps in the range of $0.1$ to $4.5$ eV (solar irradiation range), we retrain the AvgNorm FE GNN model for predicting $\alpha$ as before, but now with two additional neurons in the output layer for predicting the short-circuit current (J$_\text{sc}$) and the logarithm of the reverse saturation current (log(J$_{0}$)). The loss function is further regularized with two additional loss terms for both J$_\text{sc}$ and log(J$_{0}$). By training the model using all OPT and $80\;\%$ of the MBJ data, we evaluate its performance on the remaining MBJ data, as shown in Figure \ref{fig:fig7}. Utilizing the values of J$_\text{sc}$ and log(J$_{0}$), we calculate SLME following the procedure proposed by K. Choudhary \textit{et al.} to maximize the power density output from a solar cell \cite{choudhary2019accelerated}
\begin{equation}
\text{SLME}=\frac{\max \left\{\left(J_{\text {sc }}-J_0\left(\mathrm{e}^{e V / k T}-1\right)\right) V\right\}_V}{\int_0^{\infty} E I_{\text {sun }}(E) \mathrm{d} E}
\end{equation}
where \( J_{\text{sc}} = e \int_0^{\infty} a(E) I_{\text{sun}}(E) \, \mathrm{d}E \) with \( I_{\text{sun}} \) representing the AM1.5G solar spectrum \cite{choudhary2019accelerated, collins1972backward}. The absorbance, $a(E)$, is determined from the absorption coefficient and the film thickness ($L$) as \( a(E) = 1 - \mathrm{exp} ({-2 (2 \sqrt{2} \pi (E/hc) \sqrt{\alpha (E)}) L}) \). \( J_{0} = e \pi \int_0^{\infty} a(E) I_{\text{bb}}(E, T) \, \mathrm{d}E \) accounts for the radiative component of the electron-hole recombination current at equilibrium in darkness, with \( I_{\text{bb}} \) signifying the blackbody irradiation. We assume thin films with a thickness of $50$ nm operating at $300$ K for all materials.

\begin{table*}[ht]
\centering
\caption{\textbf{Comparison of various GNN and non-graph-based ML models for predicting solar absorption efficiency parameters:} Short-circuit current (J$_\text{sc}$), natural logarithm of the reverse saturation current (log(J$_0$)), and the spectroscopic limit of maximum efficiency (SLME)—against test set DFT (MBJ) values. The GNN models differ in their learning approach: they are either estimating the solar parameters from a learned frequency-dependent absorption coefficient, learning the solar parameters directly, or applying a learning bias to these parameters while concurrently learning the absorption coefficient. J$_\text{sc}$ and J$_0$ are measured in amperes.}

\label{tab:tab1}
\begin{tabularx}{\textwidth}{>{\hsize=1.5\hsize}X>{\hsize=0.83\hsize\centering\arraybackslash}X>{\hsize=0.83\hsize\centering\arraybackslash}X>{\hsize=0.83\hsize\centering\arraybackslash}X}
\hline
\textbf{Model (Learnables)} & \textbf{J$_\text{sc}$ (R$^2$/MAE)} & \textbf{log(J$_{0}$) (R$^2$/MAE)} & \textbf{SLME (R$^2$/MAE)} \\
\hline
GNN ($\alpha(E)$) & 0.87/33.00 & -1.86/60.26 & -0.15/5.04 \\
GNN ($\{\text{J}_{\text{sc}}, \log(\text{J}_0)\}$) & 0.87/25.87 & 0.81/11.63 & 0.70/2.18 \\
GNN ($\alpha(E)$ + $\{\text{J}_{\text{sc}}, \log(\text{J}_0)\}$ bias) & \textbf{0.90/22.42} & \textbf{0.87/10.05} & \textbf{0.76/2.04} \\
Random Forest ($\{\text{J}_{\text{sc}}, \log(\text{J}_0)\}$) & 0.71/50.59 & 0.69/17.60 & 0.60/3.10 \\
XGBoost ($\{\text{J}_{\text{sc}}, \log(\text{J}_0)\}$) & 0.71/46.95 & 0.71/15.96 & 0.59/2.93 \\
LightGBM ($\{\text{J}_{\text{sc}}, \log(\text{J}_0)\}$) & 0.75/44.18 & 0.74/15.18 & 0.63/2.84 \\
\hline
\end{tabularx}
\end{table*}

The efficacy of the GNN model in forecasting solar efficiency parameters is demonstrated in Figure \ref{fig:fig7}, where it achieves a MAE of $2.04$ and a coefficient of determination (R$^2$) exceeding $0.75$ for SLME prediction on unseen materials. This performance demonstrates superiority over traditional non-graph-based ML models. To benchmark this, we trained several non-graph-based ML models, including random forests from scikit-Learn \cite{pedregosa2011scikit} and gradient boosting decision trees from XGBoost and LightGBM packages \cite{chen2016xgboost, ke2017lightgbm}, on the MBJ dataset using features derived from chemical composition and crystal structure provided by the automatminer package \cite{dunn2020benchmarking}. The LightGBM model achieves the best performance on the test set among non-graph-based ML models, as outlined in Table \ref{tab:tab1}. Compared to the performance metrics of the solar-biased GNN model presented in Figure \ref{fig:fig7} and denoted as GNN ($\alpha(E)$ + $\{\text{J}_{\text{sc}}, \log(\text{J}_0)\}$ bias) in Table \ref{tab:tab1}, the GNN model demonstrates superior efficacy. This is manifested by the MAE for the best non-graph-based LightGBM model being approximately $97\%$, $51\%$, and $39\%$ higher for J$_\text{sc}$, log(J$_{0}$), and SLME, respectively, compared to the solar-biased GNN model. Further details about the GNNs, as well as the features and hyperparameters used for the non-graph-based ML models, are provided in the SI.  

Furthermore, a GNN model with the same architecture, yet with an output layer of only two neurons and a loss function tailored exclusively for predicting J$_\text{sc}$ and log(J$_{0}$), denoted by GNN ($\{\text{J}_{\text{sc}}, \log(\text{J}_0)\}$) in Table \ref{tab:tab1}, yields inferior performance compared to the solar-biased GNN, with the MAE higher by approximately $15\%$, $16\%$, and $7\%$ for J$_\text{sc}$, log(J$_{0}$), and SLME, respectively. Thus, we can notice that leveraging a multi-output GNN initially designed for learning the absorption spectrum, when further refined with a learning bias to emphasize solar parameters' learning, can yield improved predictive performance in forecasting solar parameters compared to directly learning them as singular targets. This improvement can be attributed to the fact that these solar parameters result from convolution integrals with the frequency-dependent absorption coefficient. By synergizing the learning of the absorption coefficient with the learning of these solar parameters, the latent feature space of the GNN becomes enriched with information on the absorption spectrum, thus enhancing the model's predictive accuracy regarding solar parameters. On the other hand, a solar-unbiased GNN model with the same architecture trained solely for predicting the absorption coefficient, denoted by GNN ($\alpha(E)$) in Table \ref{tab:tab1}, achieves $\text{MAE} = 0.384$ and $r = 0.948$ for the unnormalized spectrum of $\alpha$. In comparison, the solar-biased GNN model demonstrates an almost identical performance, with $\text{MAE} = 0.386$ and $r = 0.949$. However, the solar-unbiased model yields suboptimal results for the solar parameters, as outlined in Table \ref{tab:tab1}. The MAE increases by $47\%$, $50\%$, and $147\%$ for J$_\text{sc}$, log(J$_{0}$), and SLME, respectively, compared to the solar-biased model. This suggests that incorporating a learning bias via simple regularization terms in the loss function, aimed at emphasizing specific practical physical properties, can aid in distributing the error of the multi-output prediction in a way that substantially enhances the learning of these physical properties without compromising the overall accuracy of predicting the absorption spectrum.

\section{CONCLUSION}

We have developed multi-output GNN models capable of predicting the frequency-dependent imaginary dielectric function and absorption coefficient with accuracy on par with meta-GGA DFT. This enables the derivation of the complete dielectric response of any arbitrary material using only its atomic structure as input. We considered a spectrum spanning a $12$ eV range—from IR to UV—yet the employed GNN formalism offers easily adaptable spectrum ranges. We investigated the effect of spectrum scaling on the formation of the latent feature space and the GNNs' predictive capacity by comparing various scaling schemes, including UnNorm, MaxNorm, and AvgNorm models. Our findings highlight that the AvgNorm GNN model, treating the optical spectrum of any material as an average norm and a probability distribution function, along with a softmax activation and an evenly-weighted-distribution loss function, exhibits superior performance. Furthermore, our GNN models integrate multi-fidelity learning schemes, such as transfer learning and fidelity embedding, and utilize the whole low-fidelity OPT and high-fidelity MBJ optical spectral data available in the JARVIS DFT database. We observe a notable decrease across all error metrics and an increase in correlation with the DFT (MBJ) spectra upon the incorporation of the lower-fidelity OPT data, with the fidelity embedding approach achieving moderately superior accuracy to transfer learning attributed to its broader multi-fidelity optimization scope.

We also demonstrated that the prediction of the frequency-dependent absorption coefficient by GNNs can be fine-tuned to emphasize specific scalar optoelectronic properties without compromising the overall spectrum multi-output prediction, through simple learning biases applicable to any property of interest. As a proof of concept, we optimized the learning of solar cell performance parameters—short-circuit current, reverse saturation current, and the corresponding solar cell efficiency SLME—showing that integrating these properties into the multi-output learning of the absorption coefficient leads to enhanced prediction of solar properties compared to learning them separately. This synergy of multi-output and multi-fidelity learning, along with the flexibility to apply target-specific learning biases, presents a versatile tool for the rapid screening of solid materials for a wide range of frequency-dependent optical device applications involving metals, semiconductors, or insulators.
\\

\section{Supporting Information}
To enhance reproducibility, the datasets and codes for model design and result analysis are available on GitHub: https://github.com/UMBC-STEAM-LAB/GNN-frequency-dependent-optical-properties.

\section{Acknowledgements}
The authors acknowledge the fund from the National Science Foundation (NSF) under grant number NSF DMR$-2213398$ and the Department of Energy (DOE) under grant number DE-SC$0024236$.

\section{Notes}
The authors declare no competing interests.

\section{REFERENCES}
\bibliography{Main}

\begin{figure*} 
    \centering
    \includegraphics[width=0.96\textwidth]{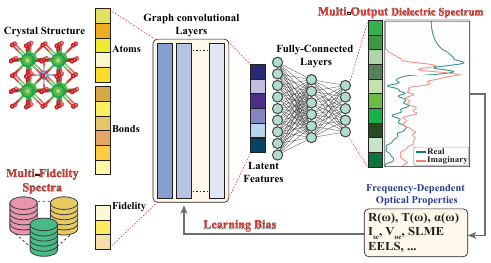}
    \caption*{\textbf{TOC Graphic.}}
    \label{fig:toc}
\end{figure*}

\end{document}


\maketitle

\newpage
\textbf{Dataset Preparation.}
The preliminary preprocessing involved refining the dataset to exclude electronically unconverged calculations, structures with band gaps $\geq 12$ eV, structures exhibiting non-positive amplitudes of the imaginary dielectric function within the 0-12 eV range, and structures characterized by extremely low density ($< 0.005$ atom/\AA{}$^3$). 

In addition, the complex shift (CSHIFT) employed by VASP in optical spectra calculations leads to smoothing for both the real and imaginary parts of the dielectric tensor, resulting in an earlier onset in both the imaginary part and the absorption spectrum. To avoid this unphysical effect, we set the imaginary part and absorption coefficient to zero below the band gap value.

\textbf{MEGNet Graph convolutional layers.}
Within a MEGNet layer, the attributes of each bond (edge) are initially updated, incorporating information from the bond itself, its connected atoms (vertices), and the global state vector, as follows 
\begin{equation}
\mathbf{e}_{ij}^{\prime}=\phi_{\mathbf{e}}(\mathbf{e}_{ij} \oplus (\mathbf{v}_{i}, \mathbf{v}_{j}) \oplus \mathbf{u})
\end{equation}
where $\oplus$ is a concatenation operator. Subsequently, the attributes of each atom (vertex) are updated, considering data from the atom itself, the average of bonds connected to it, and the global state vector. 
\begin{equation}
\mathbf{v}_{i}^{\prime}=\phi_{\mathbf{v}}(\mathbf{v}_{i} \oplus \overline{\mathbf{e}_{ij}^{\prime}} \oplus \mathbf{u})
\end{equation}
Finally, the global state attributes are updated using information from the global state itself, as well as the averages of all atoms and bonds in the structure. 
\begin{equation}
\mathbf{u}^{\prime}=\phi_{\mathbf{u}}(\mathbf{u} \oplus \overline{\mathbf{v}^{\prime}}
\oplus \overline{\mathbf{e}^{\prime}}) 
\end{equation}
A set of dense layers is utilized for the update functions $(\phi_{\mathbf{e}}, \phi_{\mathbf{v}}, \phi_{\mathbf{u}})$.

\textbf{Hyperparameter Optimization.} The following hyperparameters were explored within the specified search spaces using Bayesian multi-objective optimization from Optuna, with a total of $200$ trials: \\ 
- Number of atom features = (12, 30, step=2) \\
- Number of bond features = (120, 200, step=20) \\
- Dropout rate = (0.00, 0.20, step=0.05) \\
- Batch size = (80, 240, step=20)  \\
- Dimensions of hidden layers: $n_1$ = (300, 600, step=25), \\
$n_2$ = (200, 400, step=25), $n_3$ = (100, 200, step=25), $n_4$ = (25, 200, step=25), \\ 
$n_5$ = (25, 100, step=25), $n_6$ = (25, 200, step=25), $n_7$ = (100, 400, step=25) 

\begin{figure*}
    \centering
    \includegraphics[width=0.84\textwidth]{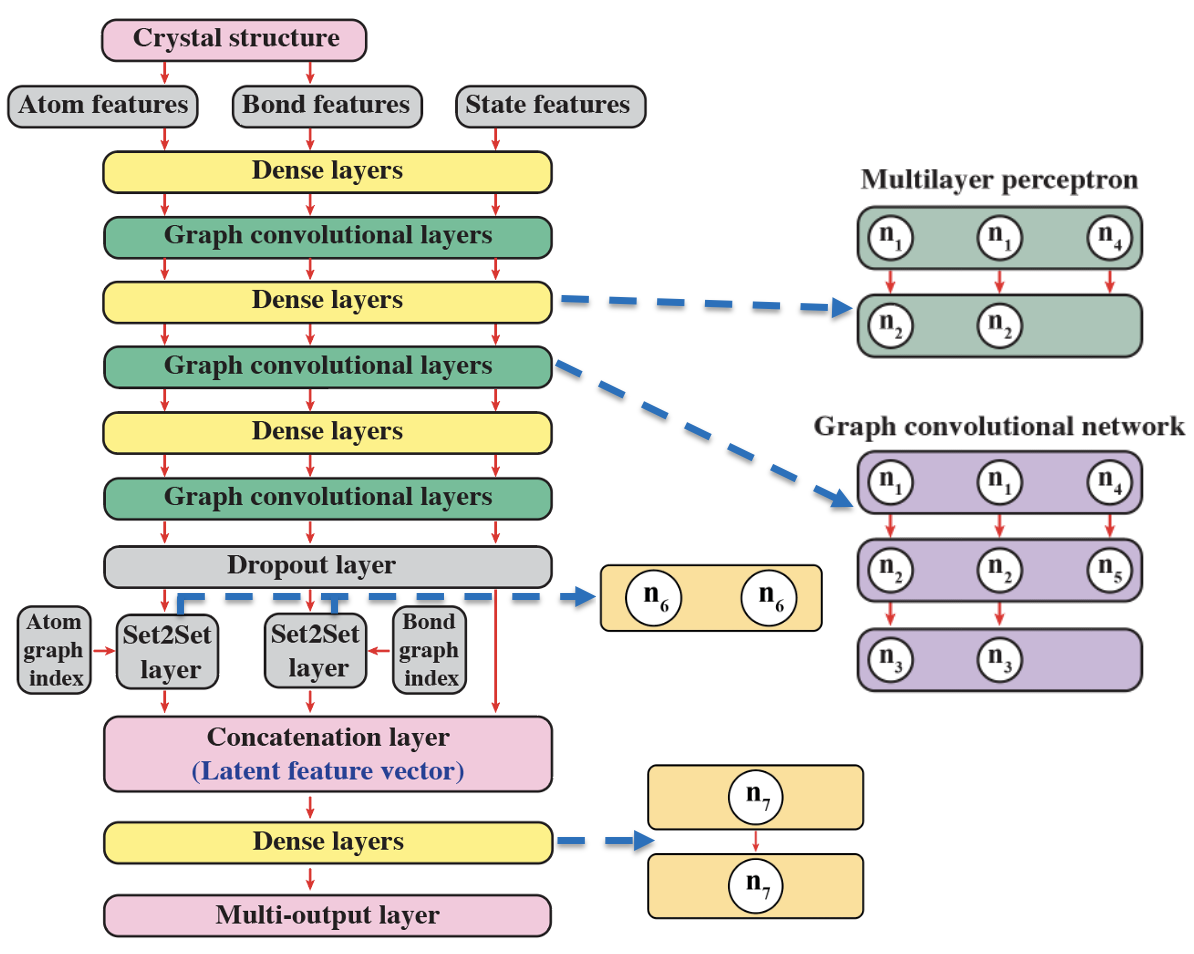}
    \caption{A detailed overview of the GNN model architecture.}
    \label{fig:fig_s1}
\end{figure*}

The set of hyperparameters yielding the best performance on the validation set comprises: 22 atom features, 180 bond features, a dropout rate of 0.05, a batch size of 110, and layer dimensions of $n_1$ = 400, $n_2$ = 300, $n_3$ = 150, $n_4$ = 150, $n_5$ = 75, $n_6$ = 75, and $n_7$ = 400. The internal neuron structure of all dense layers, graph convolutional layers, and the set2set layer is detailed in Figure \ref{fig:fig_s1}. The pre-processing dense layers for state features use $n_4$ = 150 neurons, except the initial pre-processing layer in single-fidelity models, which includes a non-trainable $2$-neuron layer to transmit dummy $2$ placeholder state features to the first graph convolutional layer. Other hyperparameters were determined heuristically, including the number of dense/graph convolutional layers, the Gaussian width and cutoff radius for bond features, and the initial learning rate ($lr$). 
The Exponential Linear Unit (ELU) activation function was utilized for all neurons in the pre-processing dense, graph convolutional, and post-processing layers. The output layers employed either linear or Rectified Linear Unit activation (ReLU) based on the optical spectrum scaling architecture as discussed in the "Multi-Output Spectrum Architecture" section.

The model training process included a maximum of $320$ epochs on the training dataset, implementing an early stopping criterion to halt the process upon achieving convergence in the validation loss to mitigate overfitting. The weight optimization employed the AdamW optimizer.

\newpage \textbf{Effect of Multi-Output Spectrum Architecture on GNN Performance.} The median and interquartile range (IQR) of error distributions for various metrics, calculated using single-fidelity (SF) GNNs trained solely on the MBJ dataset with various multi-output spectrum architectures to predict the imaginary part of the dielectric function ($\varepsilon_2$) and the reduced absorption coefficient ($\alpha$), are presented in Table \ref{tab:tabs1} and Table \ref{tab:tabs2}, respectively. 
\begin{table*}
\centering
\caption{Numerical Representation, median (IQR), of GNN performance metrics on $\varepsilon_2$ in Figure 3-a).}
\begin{adjustbox}{width=0.85\textwidth}
\begin{tabularx}{\linewidth}{l *{5}{>{\centering\arraybackslash}X}}
    \hline
    Property & Metric & $\varepsilon_2^{\text{UnNorm (MAE)}}$ & $\varepsilon_2^{\text{MaxNorm (MAE)}}$ & $\varepsilon_2^{\text{AvgNorm (KL)}}$ & $\varepsilon_2^{\text{AvgNorm (MAE)}}$ \\
    \hline
    $\varepsilon_2$ & MAE & 1.02 (1.71) & 1.25 (2.02) & 1.10 (1.78) & 0.915 (1.71) \\ 
    $r$ & Pearson & 0.940 (0.171) & 0.939 (0.152) & 0.945 (0.142) & 0.954 (0.133) \\ 
    $\varepsilon_2^{\text{max}}$ & MAE & 3.62 (18.1) & 4.05 (19.5) & 3.28 (19.0) & 2.88 (17.4) \\ 
    $\varepsilon_2^{\text{avg}}$ & MAE & 0.329 (0.802) & 0.469 (1.03) & 0.249 (0.814) & 0.260 (0.807) \\ 
    $\overline{\varepsilon}_2$ ($\times 10^3$) & MAE & 0.764 (0.826) & 0.790 (0.809) & 0.740 (0.739) & 0.663 (0.731) \\ 
    $\overline{\varepsilon}_2$ ($\times 10^3$) & KL & 0.083 (0.189) & 0.058 (0.124) & 0.050 (0.102) & 0.041 (0.090) \\  
    $\overline{\varepsilon}_2$ ($\times 10^3$) & WD & 6.63 (10.4) & 6.83 (10.6) & 6.04 (8.62) & 5.23 (8.25) \\  
    $\overline{\varepsilon}_2'$ ($\times 10^3$) & MAE & 2.88 (2.35) & 3.24 (2.36) & 2.77 (2.00) & 2.61 (2.07)  \\  
    moments & MAE & 0.181 (0.331) & 0.196 (0.363) & 0.162 (0.288) & 0.142 (0.267) \\
    \hline
\end{tabularx}
\end{adjustbox}
\label{tab:tabs1}
\end{table*}

\begin{table*}
\centering
\caption{Numerical Representation, median (IQR), of GNN performance metrics on $\alpha$ in Figure 3-a).}
\begin{adjustbox}{width=0.85\textwidth}
\begin{tabularx}{\linewidth}{l *{5}{>{\centering\arraybackslash}X}}
    \hline
    Property & Metric & $\alpha^{\text{UnNorm (MAE)}}$ & $\alpha^{\text{MaxNorm (MAE)}}$ & $\alpha^{\text{AvgNorm (KL)}}$ & $\alpha^{\text{AvgNorm (MAE)}}$ \\
    \hline
    $\alpha$ & MAE & 0.969 (1.80) & 1.16 (2.24) & 1.09 (2.06) & 0.891 (1.85) \\  
    $r$ & Pearson & 0.912 (0.228) & 0.909 (0.234) & 0.915 (0.190) & 0.929 (0.206) \\ 
    $\alpha^{\text{max}}$ & MAE & 2.67 (11.1) & 2.82 (11.1) & 2.40 (12.3) & 2.16 (10.7) \\ 
    $\alpha^{\text{avg}}$ & MAE & 0.294 (0.850) &  0.452 (1.06) & 0.271 (0.885) & 0.261 (0.849) \\ 
    $\overline{\alpha}$ ($\times 10^3$) & MAE & 0.881 (1.01) & 0.901 (1.00) & 0.873 (0.840) & 0.793 (0.931) \\ 
    $\overline{\alpha}$ ($\times 10^3$) & KL & 0.100 (0.296) & 0.072 (0.180) & 0.061 (0.136) & 0.057 (0.146) \\  
    $\overline{\alpha}$ ($\times 10^3$) & WD & 7.68 (12.2) & 8.34 (13.0) & 7.36 (10.7) & 6.54 (11.3) \\  
    $\overline{\alpha}'$ ($\times 10^3$) & MAE & 3.67 (3.61) & 3.73 (3.14) & 3.49 (2.54) & 2.99 (2.61)  \\  
    moments & MAE & 0.207 (0.411) & 0.213 (0.444) & 0.195 (0.382) & 0.174 (0.360) \\
    \hline
\end{tabularx}
\end{adjustbox}
\label{tab:tabs2}
\end{table*}

Additional hyperparameters include $lr$ and weight for the normalized spectrum in the loss function, formulated as $\mathcal{L} = \mathcal{L}_N + w \mathcal{L}_{S}$, where $\mathcal{L}_{S}$ represents the loss in the normalized spectrum ($\overline{\varepsilon}_2$ or $\overline{\alpha}$). For $\varepsilon_2$: $lr = 1 \times 10^{-3}$ for UnNorm; $lr = 1 \times 10^{-6}$, $w = 1.5 \times 10^3$ for MaxNorm (MAE); $lr = 1 \times 10^{-3}$, $w = 7.3$ for AvgNorm (KL); and $lr = 1 \times 10^{-3}$, $w = 4.6 \times 10^3$ for AvgNorm (MAE). For $\alpha$: $lr = 7 \times 10^{-4}$ for UnNorm; $lr = 1 \times 10^{-6}$, $w = 9.5 \times 10^2$ for MaxNorm (MAE); $lr = 1 \times 10^{-3}$, $w = 7.0$ for AvgNorm (KL); and $lr = 1 \times 10^{-3}$, $w = 3.7 \times 10^3$ for AvgNorm (MAE).

\begin{figure*} 
    \centering
    \includegraphics[width=0.96\textwidth]{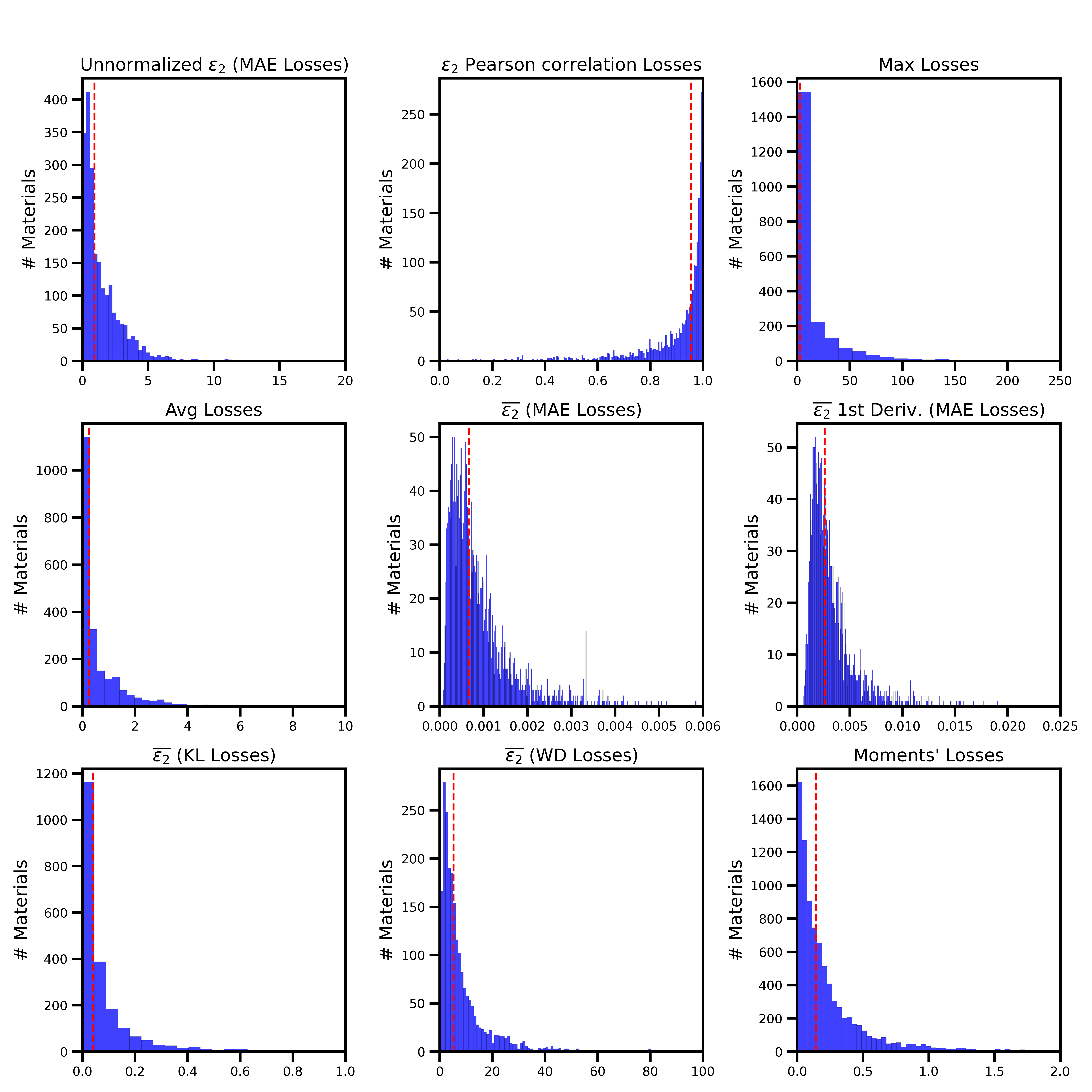}
    \caption{Error distribution for the imaginary component of the dielectric function ($\varepsilon_2$) predicted for the MBJ test set by the SF AvgNorm (MAE) GNN model. The dashed red lines represent the median values.} 
     \label{fig:fig_s2}
\end{figure*}

\begin{figure*} 
    \centering
    \includegraphics[width=0.96\textwidth]{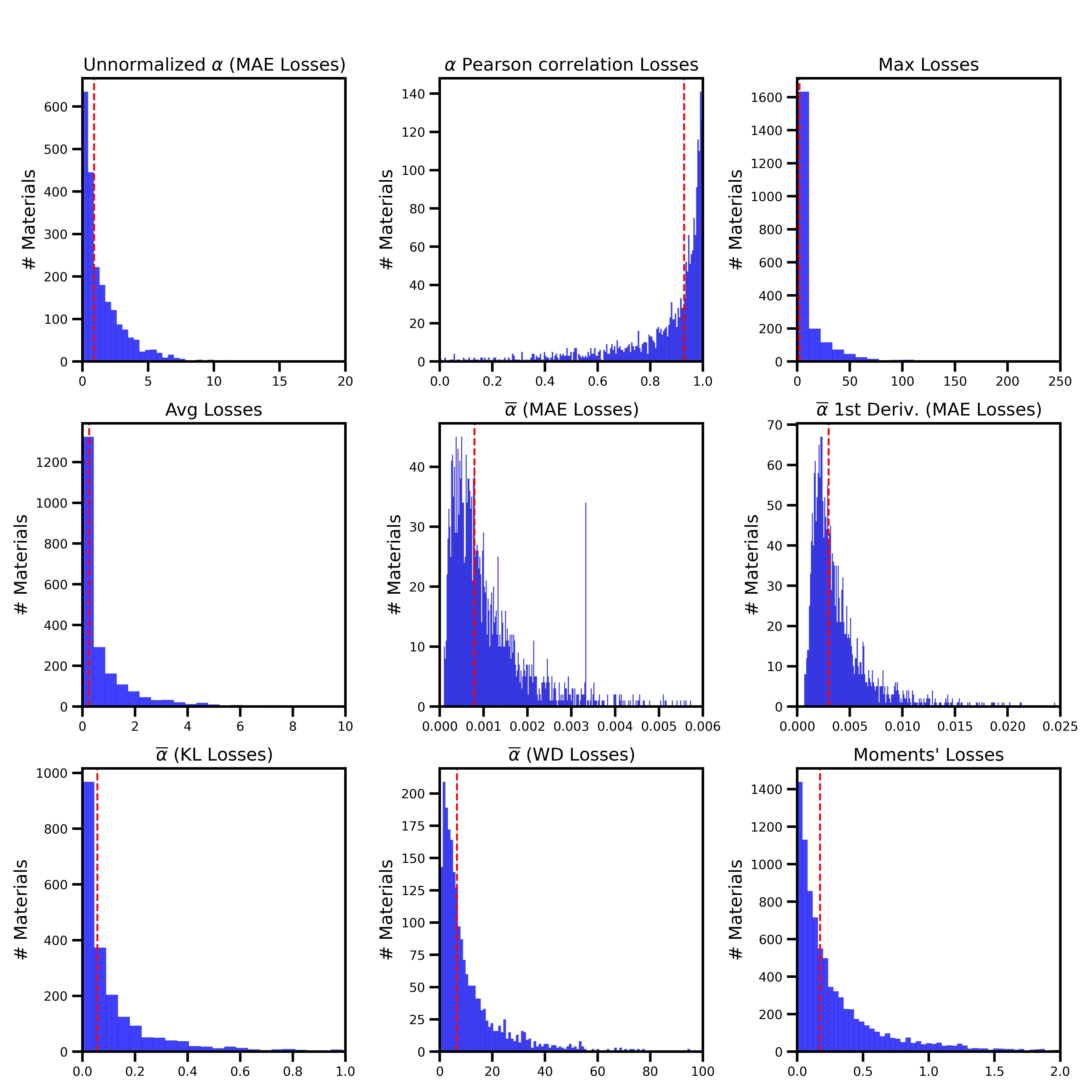}
    \caption{Error distribution for the reduced absorption coefficient ($\alpha$) predicted for the MBJ test set by the SF AvgNorm (MAE) GNN model. The dashed red lines represent the median values.} 
     \label{fig:fig_s3}
\end{figure*}

\newpage
\textbf{Effect of Multi-Fidelity Learning on GNN Performance.} The median and IQR of error distributions for various metrics, calculated using the optimal SF GNN model (trained solely on MBJ data) and multi-fidelity GNNs, employing transfer learning (TL) and fidelity-embedding (FE) strategies (trained on both MBJ and OPT datasets), are outlined for $\varepsilon_2$ and $\alpha$ in Tables \ref{tab:tabs3} and \ref{tab:tabs4}, respectively.

\begin{table*}
\centering
\caption{Numerical Representation, median (IQR), of GNN performance metrics on $\varepsilon_2$ in Figure 4).}
\begin{adjustbox}{width=0.85\textwidth}
\begin{tabularx}{\linewidth}{l *{4}{>{\centering\arraybackslash}X}} 
    \hline
    Property & Metric & {\text{SF}}& {\text{TL}} & {\text{FE}} \\
    \hline
    $\varepsilon_2$ & MAE & 0.915 (1.71) & 0.709 (0.876) & 0.690 (0.877) \\  
    $r$ & Pearson & 0.954 (0.133) & 0.976 (0.058) & 0.978 (0.056) \\ 
    $\varepsilon_2^{\text{max}}$ & MAE & 2.88 (17.4) & 2.00 (9.52) & 1.96 (8.74) \\ 
    $\varepsilon_2^{\text{avg}}$ & MAE & 0.260 (0.807) & 0.181 (0.406) & 0.173 (0.377) \\ 
    $\overline{\varepsilon}_2$ ($\times 10^3$) & MAE & 0.663 (0.731) & 0.464 (0.460) & 0.449 (0.472) \\ 
    $\overline{\varepsilon}_2$ ($\times 10^3$) & KL & 0.041 (0.090) & 0.020 (0.043) & 0.022 (0.046) \\  
    $\overline{\varepsilon}_2$ ($\times 10^3$) & WD & 5.23 (8.25) & 3.54 (5.08) & 3.24 (4.54) \\  
    $\overline{\varepsilon}_2'$ ($\times 10^3$) & MAE & 2.61 (2.07) & 2.39 (1.70) & 2.00 (1.60) \\  
    moments & MAE & 0.142 (0.267) & 0.103 (0.171) & 0.090 (0.156) \\
    \hline
\end{tabularx}
\end{adjustbox}
\label{tab:tabs3}
\end{table*}

\begin{table*}
\centering
\caption{Numerical Representation, median (IQR), of GNN performance metrics on $\alpha$ in Figure 4).}
\begin{adjustbox}{width=0.85\textwidth}
\begin{tabularx}{\linewidth}{l *{4}{>{\centering\arraybackslash}X}} 
    \hline
    Property & Metric & {\text{SF}}& {\text{TL}} & {\text{FE}} \\
    \hline
    $\alpha$ & MAE & 0.891 (1.85) & 0.713 (0.994) & 0.668 (0.965) \\ 
    $r$ & Pearson & 0.929 (0.206) & 0.963 (0.078) & 0.965 (0.080) \\ 
    $\alpha^{\text{max}}$ & MAE & 2.16 (10.7) & 1.64 (5.81) & 1.60 (5.98) \\ 
    $\alpha^{\text{avg}}$ & MAE & 0.261 (0.849) & 0.193 (0.422) & 0.176 (0.387) \\ 
    $\overline{\alpha}$ ($\times 10^3$) & MAE & 0.793 (0.931) & 0.561 (0.604) & 0.541 (0.583) \\ 
    $\overline{\alpha}$ ($\times 10^3$) & KL & 0.057 (0.146) & 0.029 (0.073) & 0.028 (0.068) \\  
    $\overline{\alpha}$ ($\times 10^3$) & WD & 6.54 (11.3) & 4.26 (6.37) & 3.93 (5.72) \\  
    $\overline{\alpha}_2'$ ($\times 10^3$) & MAE & 2.99 (2.61) & 3.50 (2.85) & 2.45 (2.12) \\  
    moments & MAE & 0.174 (0.360) & 0.122 (0.220) & 0.103 (0.197) \\
    \hline
\end{tabularx}
\end{adjustbox}
\label{tab:tabs4}
\end{table*}

For TL on $\varepsilon_2$: $w = 3.8 \times 10^{3}$, $lr = 7 \times 10^{-4}$ for the initial training on all OPT data; and $w = 4.6 \times 10^{3}$, $lr = 7 \times 10^{-4}$ for training the dense layers post-LFV on MBJ data. For TL on $\alpha$: $w = 3.0 \times 10^{3}$, $lr = 3 \times 10^{-4}$ for the initial training on all OPT data; and $w = 3.7 \times 10^{3}$, $lr = 3 \times 10^{-4}$ for training the dense layers post-LFV on MBJ data. Additionally, we increased the number of post-processing dense layers following the LFV from two to three, with each layer comprising $1000$ neurons instead of $n_7 = 400$.

For FE on $\varepsilon_2$: length of the FE state vector $= 20$, $lr = 7 \times 10^{-4}$, and $w = 4.3 \times 10^3$. For FE on $\alpha$, length of the FE state vector $= 16$, $lr = 7 \times 10^{-4}$, and $w = 3.7 \times 10^3$.

\begin{figure*} 
    \centering
    \includegraphics[width=0.96\textwidth]{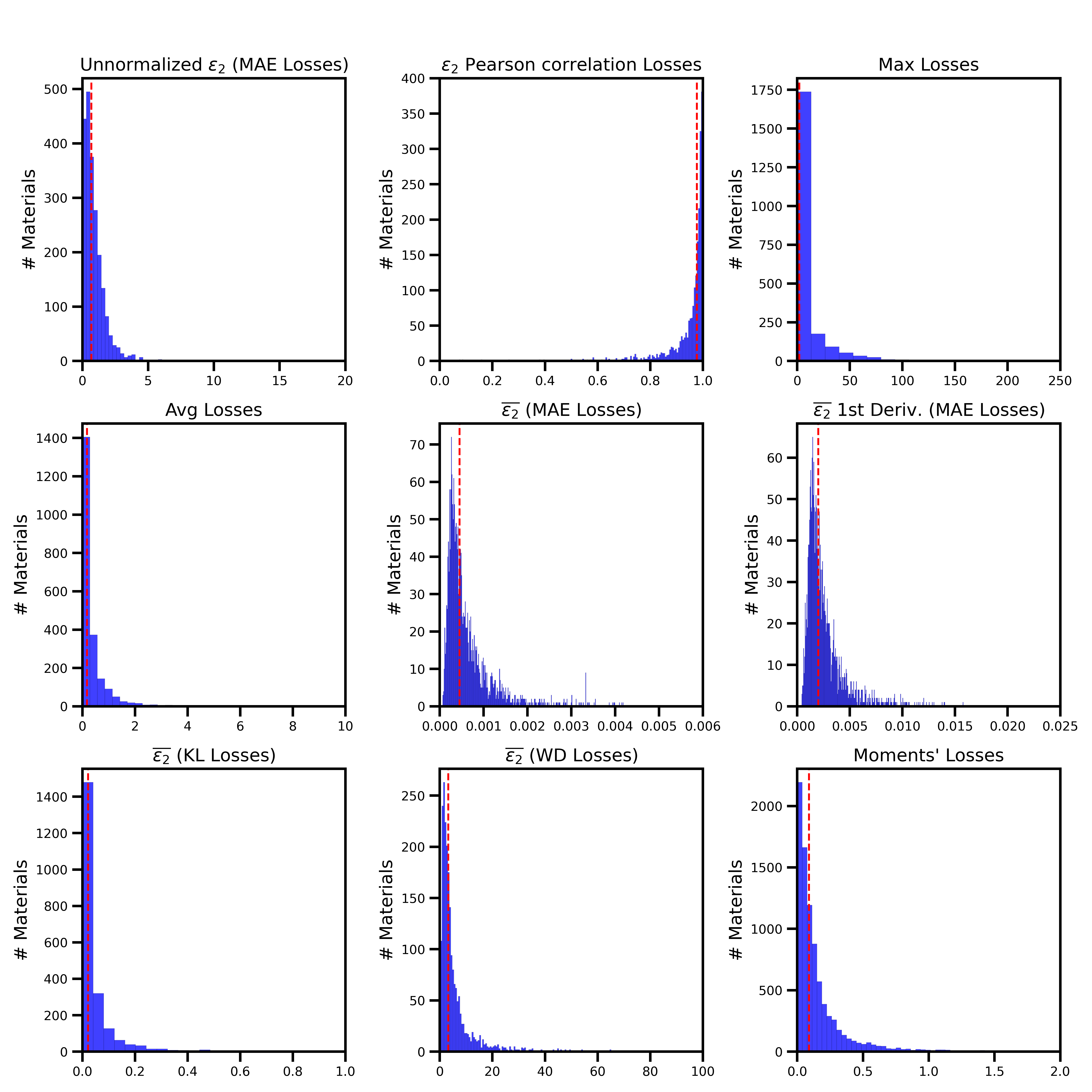}
    \caption{Error distribution for the imaginary component of the dielectric function ($\varepsilon_2$) predicted for the MBJ test set by the FE AvgNorm (MAE) GNN model. The dashed red lines represent the median values.} 
     \label{fig:fig_s4}
\end{figure*}

\begin{figure*} 
    \centering
    \includegraphics[width=0.96\textwidth]{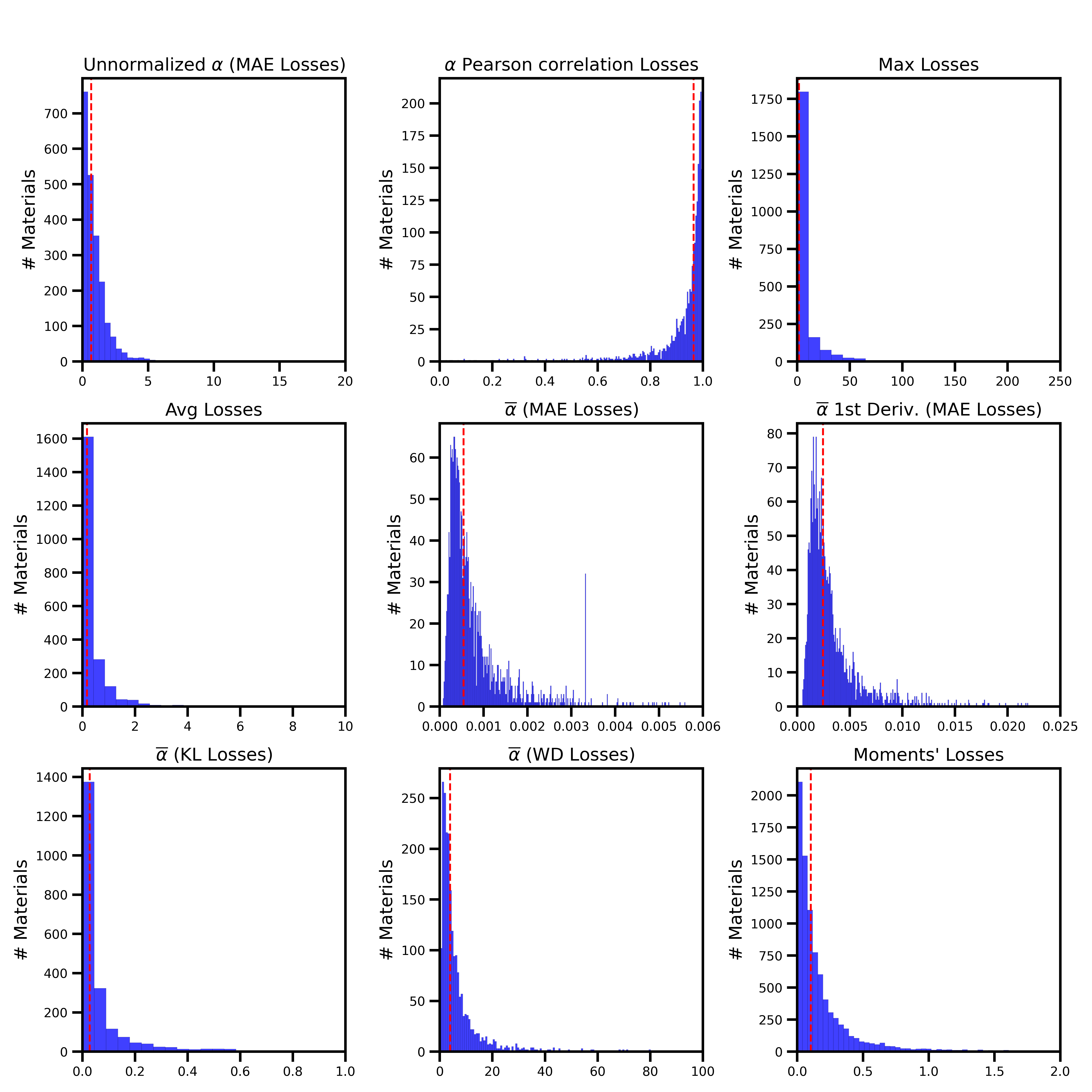}
    \caption{Error distribution for the reduced absorption coefficient ($\alpha$) predicted for the MBJ test set by the FE AvgNorm (MAE) GNN model. The dashed red lines represent the median values.} 
     \label{fig:fig_s5}
\end{figure*}

\newpage
\textbf{Feature Correlation with Multi-Output Optical Spectrum Properties for Different GNN Models.}
\begin{figure*}
    \centering
    \includegraphics[width=0.62\textwidth]{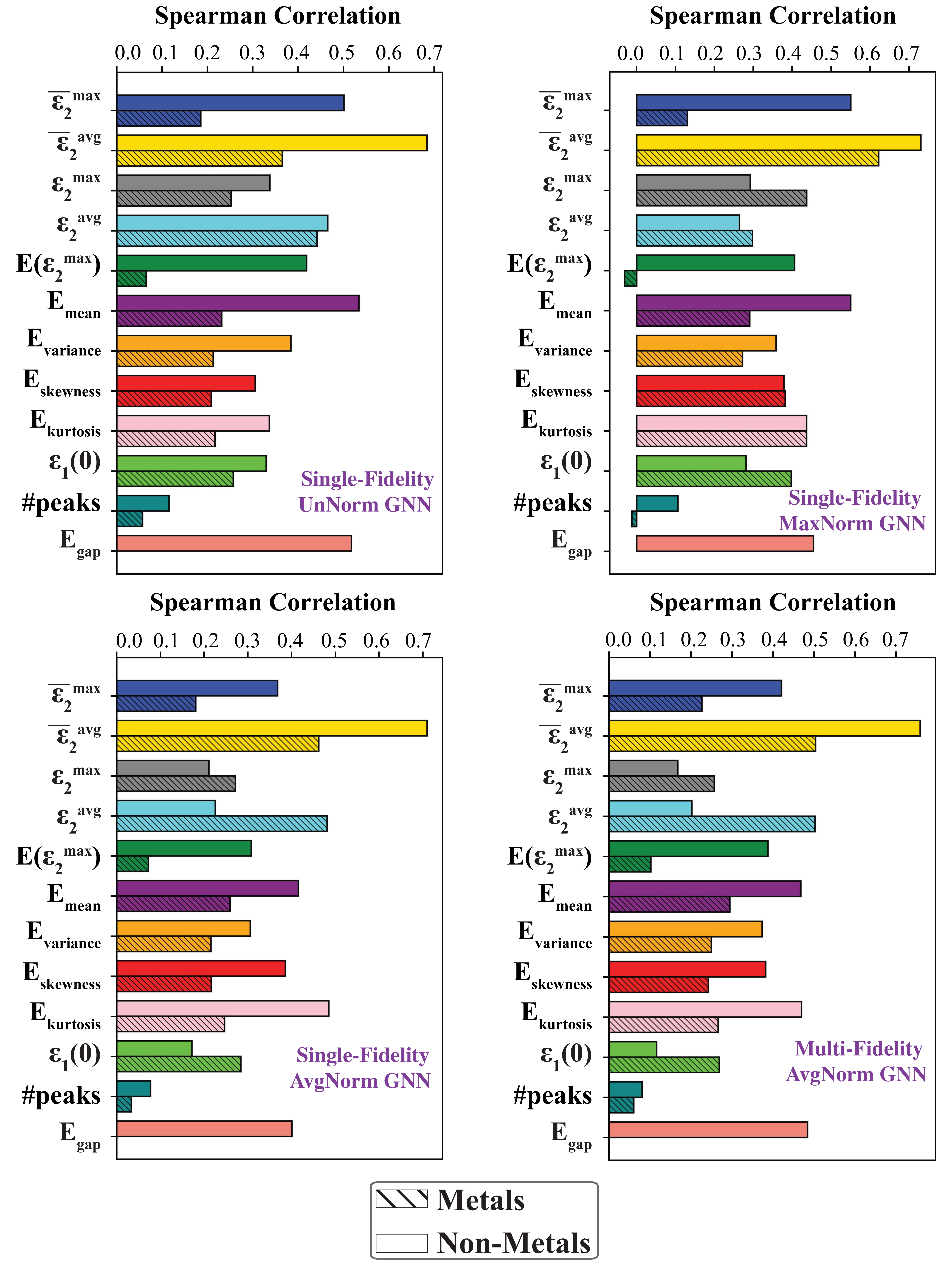}
    \caption{Spearman rank-order correlation coefficients between the normalized Euclidean pairwise distances of latent feature vectors and various scalar/vector properties extracted from the optical spectrum are computed for both metals and non-metals within the MBJ dataset. The correlations are presented for latent features from the UnNorm, MaxNorm, and AvgNorm (MAE) single-fidelity models, each exclusively trained using the MBJ dataset. Additionally, results from the multi-fidelity FE model (shown in Figure 5-b) in the manuscript), which trains jointly on OPT/MBJ data, are included.}
    \label{fig:fig_s6}
\end{figure*}

\newpage
\textbf{Predicting Solar Absorption Efficiency with GNNs and Traditional Machine Learning Models.} The featurizers utilized from the automatminer package include: \\
\{Composition: \text{ElementProperty.from\_preset("magpie")},\\
Structure:
\text{EwaldEnergy}, \text{SineCoulombMatrix}, \text{DensityFeatures}, \\
\text{SiteStatsFingerprint.from\_preset("CrystalNNFingerprint\_ops")}, \text{GlobalSymmetryFeatures}\}

The optimal number of trees (estimators) for the three models (Random Forest, XGBoost, LightGBM), reported in Table 1, is determined to be $500$. The loss weights for the three solar GNN models, reported in Table 1, were optimized to $[1.0, 3.4 \times 10^{3}]$ for $\alpha^{\text{avg}}$ and $\overline{\alpha}$. The same hyperparameters from the previous FE GNN were adapted, including $lr = 7 \times 10^{-4}$ and a FE state vector of length $16$. The learning bias for the solar parameters J$_\text{sc}$ and log(J$_0$) was optimized to $0.01$. Thus, the weights for the loss components of the solar-biased GNN are set to $[1.0, 3.4 \times 10^{3}, 0.01, 0.01]$ for $\alpha^{\text{avg}}$, $\overline{\alpha}$, J$_\text{sc}$, and log(J$_0$).

\begin{figure*} 
    \centering
    \includegraphics[width=0.75\textwidth]{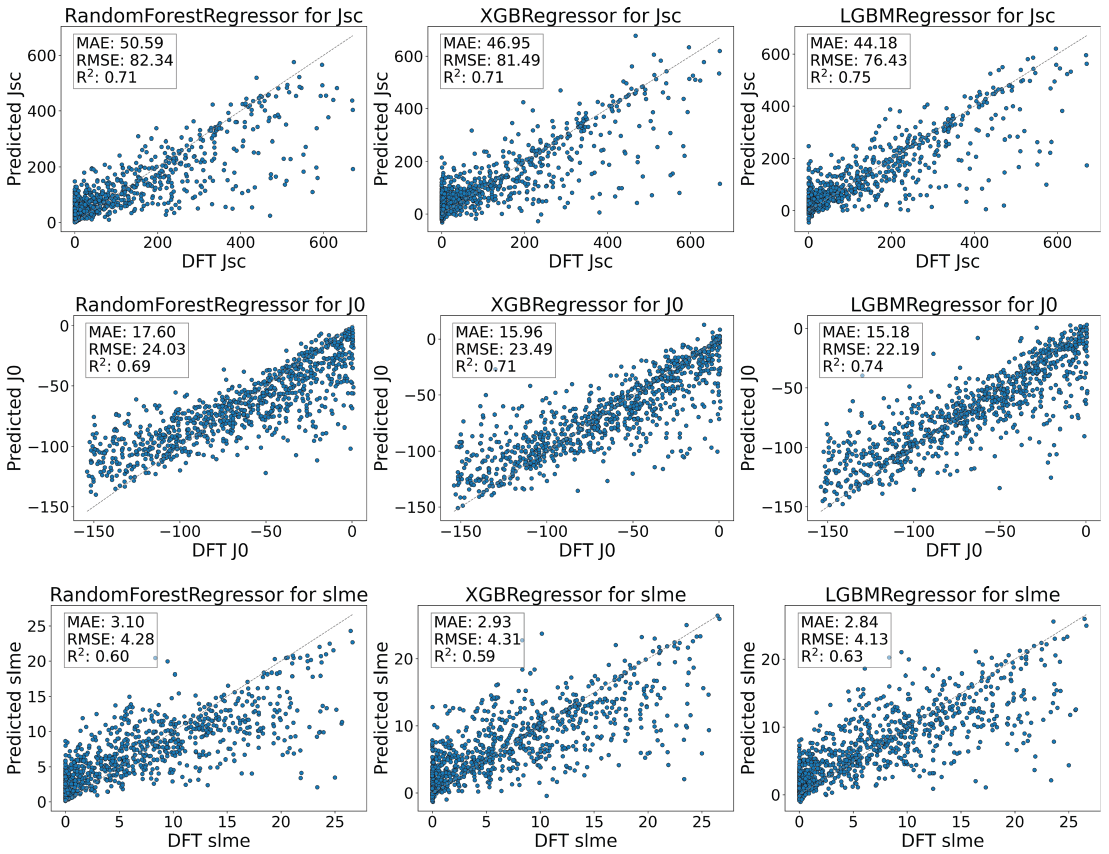}
    \caption{Performance validation of three non-graph-based machine learning models on the MBJ test set materials with band gaps spanning $0.1$ to $4.5$ eV (solar irradiation range), comparing predictions of the short-circuit current (J$_\text{sc}$), the natural logarithm of the reverse saturation current (log(J$_0$)), and the spectroscopic limit of maximum efficiency (SLME) against DFT (MBJ) calculations. J$_\text{sc}$ and J$_0$ are measured in amperes (A).}
     \label{fig:fig_s7}
\end{figure*}